\documentclass{emulateapj}

\newcommand{\HI}{\ion{H}{1}~}
\newcommand{\kms}{km~s$^{-1}~$}
\newcommand{\Lya}{Ly$\alpha$ }

\usepackage{lscape} 
\usepackage{multirow}
\usepackage{color} 
\usepackage{graphics,graphicx}

\slugcomment{Accepted for publication in the Astrophysical Journal}

\shorttitle{Discovery of a DLA Near a Galaxy Undergoing Gas Inflow}
\shortauthors{Borthakur et al.}

\begin{document}

\title{Discovery of a Damped Ly$\alpha$ System in a Low-z Galaxy Group: Possible Evidence for Gas Inflow and Nuclear Star Formation}

\author{Sanchayeeta Borthakur}
\affil{School of Earth and Space Exploration, Arizona State University, 781 Terrace Mall, Tempe, AZ
 85287, USA }
\email{sanch@asu.edu}

\author{Emmanuel Momjian}
\affil{National Radio Astronomy Observatory, P.O. Box O, Socorro, NM 87801, USA}

\author{Timothy M. Heckman}
\affil{Department of Physics \& Astronomy, Johns Hopkins University, Baltimore, MD, 21218, USA }

\author{Barbara Catinella}
\affil{International Centre for Radio Astronomy Research, The University of Western Australia, 35 Stirling Highway, Crawley WA 6009, Australia}

\author{Fr\'ed\'eric P. A. Vogt}
\affil{European Southern Observatory, Av. Alonso de Cordova 3107, 763 0355 Vitacura, Santiago, Chile}

\author{Jason Tumlinson}
\affil{ Space Telescope Science Institute, 3700 San Martin Drive, Baltimore, MD 21218, USA;
 Department of Physics \& Astronomy, Johns Hopkins University, Bloomberg Centre, 3400 N.
 Charles Str, Baltimore, MD 21218, USA}

\begin{abstract}

We present a low-redshift (z=0.029) Damped \Lya (DLA) system in the spectrum of a background quasi-stellar object (QSO). The DLA is associated with an interacting galaxy pair within a galaxy group. We detected weak \Lya emission centered at the absorption trough of the DLA. The emission was likely tracing the neutral \HI reservoir around the galaxies in the interacting pair, which scattered the \Lya generated by star formation within those galaxies. We also found that the interacting pair is enveloped by a large \HI cloud with $\rm M(HI)=2\times 10^{10}M_{\odot}$.
We discovered blueshifted 21~cm \HI emission, corresponding to M(HI)=$\rm 2\times10^{9}~M_{\odot}$, associated with  J151225.15$+$012950.4 - one of the galaxies in the interacting pair. The blueshifted \HI was tracing gas flowing into the galaxy from behind and toward us. Gas at similar blueshifted velocities was seen in the QSO sight line thus suggesting the presence of a filamentary structure of the order of 100~kpc feeding the galaxy. 
We estimated a mass inflow rate of $ \rm 2 M_{\odot}~yr^{-1}$ into the galaxy, which matches the star formation rate estimated from H$\alpha$ emission. It is likely that the inflow of enormous amounts of gas has triggered star formation in this galaxy. The sudden acquisition of cold gas may lead to a starburst in this galaxy like those commonly seen in simulations.

\end{abstract}


\section{INTRODUCTION\label{intro}}

Damped \Lya (DLA; $\rm N(HI)>2 \times 10^{20}~cm^{-2}$) systems trace the bulk of the observed neutral gas in the universe and therefore are powerful probes of galaxy formation and evolution \citep{wolfe98,prochaska09, noterdaeme09}. 
The unique property of the gas in these systems is its ability to self-shield itself from becoming photoionized by the cosmic ultraviolet background \citep[see review by][and references therein]{wolfe05}. 
Consequently, such clouds are likely to survive long enough to produce stars. 

\begin{figure*}[!h]
\includegraphics[trim=3mm 25mm 5mm 30mm,  clip=true, scale=0.85, angle=0]{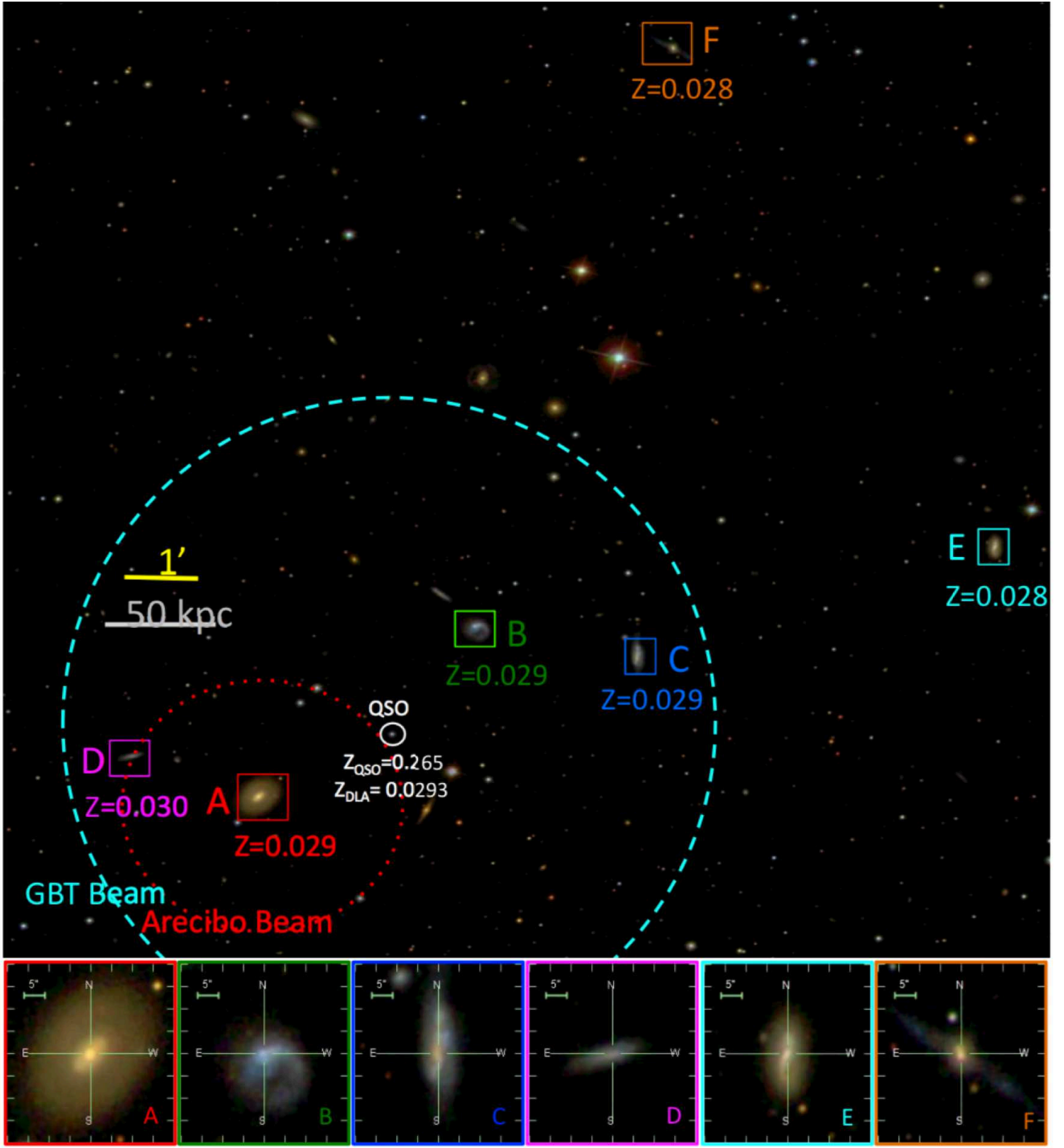}
\caption{SDSS multicolor image toward the QSO SDSS J151237.15$+$012846.0. The QSO is marked in white, and the galaxies are marked with colored squares and labeled. SDSS insets of each of the galaxies are shown below. The redshifts of the galaxies are labeled for the six spectroscopically confirmed redshifts. The red and the cyan circles represent the FWHM of the Arecibo ($4^{\prime}$) and the GBT ($9.1^{\prime}$) beams centered at their respective pointings . The VLA field of view was bigger than the image shown above. The SDSS-based group finders (Yang et al. 2007, Tago et al. 2010) listed the halo mass of this group as  $\rm 2.5\times 10^{12}~M_{\odot}$ and the virial radius as 176~kpc. Insets at rge bottom show the SDSS multicolor images of each of the galaxies. All of them were classified as blue galaxies,  with stellar masses in the range of  0.1-8.3 $\rm \times 10^{10}~M_{\odot}$ and SFRs of 0.4-5~$\rm M_{\odot}~yr^{-1}$.  }
 \label{SDSS_image}
\end{figure*}

Most of our current understanding of DLAs comes from high-redshift systems seen in the optical spectra of Quasi-stellar Objects \citep[QSOs;][and references therein]{wolfe86,lanzetta91, rao95, kulkarni97,prochaska00, nestor03, peroux03,prochaska03, khare04, turnshek04, prochaska05, ledoux06, rafelski12, rafelski14,krogager17}. 
However, a critical disadvantage of studying high-redshift DLAs is the difficulty of imaging the galaxy distribution around them to identify their host. This severely limits our understanding of the connection between DLAs and the galaxies where stars are formed. 
Therefore, low-redshift DLAs are highly valued. 
This is particularly true for DLAs at redshifts, $\rm z<0.05$, which can also be imaged in \HI 21~cm. These systems are our only resources to ascertain the nature and connection of DLAs to the gas reservoirs of galaxies. 
 It is worth noting that the origin of the low and high-redshift DLAs may different. Most of the low-redshift DLAs were believed to be associated with extended disks, the inner circumgalactic medium, or satellite galaxies \citep[e.g.][]{Neeleman17}, while simulations showed that the high-$z$ DLAs were not necessarily galaxy disks \citep{gardner01, pontzen08, yajima12,cen12, rahmati13, yuan_cen16}.

Unfortunately, at very low-redshifts, there are only a handful of DLAs that have been identified \citep{battisti12,N16}. This is because large space-based surveys are required to identify low-redshift DLAs, thus making DLAs extremely expensive to discover.
Therefore, pursuing statistical studies with large samples of low-redshift DLAs is currently almost impossible. Nevertheless, the situation is not entirely hopeless. Even individual systems promise to substantially increase our understanding and hence are worth studying.

We present one such discovery of a DLA tracing gas in a low-redshift galaxy group. This detection is the first of its kind and provided a unique window into a dynamically active environment. In fact, most galaxies in the universe reside in groups or clusters \citep{tully87}. In particular, groups are sites of intense galaxy interactions owing to the low velocity dispersion of the member galaxies of only a few hundred kilometers per second. As a result, gas-rich mergers and subsequent intense star formation are commonly seen in galaxy group environments  \citep{sanders88,sanders_miradel96}. These environments also play a crucial role in the {\it{pre-processing}} of galaxies, where galaxies lose a significant fraction of their cold-gas content and may even suffer changes in their morphology.
 Many processes may be at work during the pre-processing phase, for example, the loss of neutral gas reservoirs due to enhanced star formation or from tidal and/or ram pressure stripping \citep{fujita04,mihos04, cortese06,rasmussen08, hess13, vijayaraghavan13,cybulski14,wetzel15}.


Our current understanding of gas physics and star-formation is not robust enough to predict the resulting fate of the cold gas in galaxies involved in an interaction. Therefore, gas-rich galaxies in the early stages of interactions are perfect candidates for studying conditions that dictate the fate of the gas - whether the cold gas will trigger a starburst, become ionized, or transform into a different phase. In this paper, we studied the gas physics in this galaxy group to understand the nature, origin, survival timescales, and fate of the neutral gas  during an interaction.

This work was motivated by two main objectives: first, to explore a very low-z DLA to learn more about the DLA population, and second, to explore the properties of the intragroup medium and relate it to the cold gas within member galaxies. This work was enabled by the discovery of a DLA indicative of cold gas in a QSO sight line probing a low-redshift galaxy group.
We begin by describing the group and member galaxies in Section~2. This is followed by a detailed discussion on the absorbers detected in the QSO sight line and their properties in Section~3. In Section~4, we present our \HI 21~cm images and discuss the properties of the atomic gas in the group. 
In Section~5, we discuss the implications of our observations in the context of an interaction-driven starburst. We do so by combining the observations of different phases of gas traced by the \Lya and metal-line absorbers, as well as \HI 21~cm emission. 
Finally, we summarize our findings in Section 6.

The cosmological parameters used in this study were $H_0 =70~{\rm km~s}^{-1}~{\rm Mpc}^{-1}$, $\Omega_m = 0.3$, and $\Omega_{\Lambda} = 0.7$.

\section{The Target}%

Our target was a galaxy group that was probed by the QSO sight line SDSS~J151237.15$+$012846.0.
The galaxy group consisted of six spectroscopically confirmed galaxies at a redshift of 0.0293. 
The Sloan Digital Sky Survey Data Release 7 (SDSS DR7) group catalog published by \citet{tago10} estimated the virial radius of the group to be 176~kpc and the maximum projected linear size (including all members) to be 444~kpc.  The velocity dispersion was 200~\kms. 
Figure~\ref{SDSS_image} shows the SDSS image of the group, with the galaxies and their redshift labeled. For simplicity we named the galaxies in alphabetical order from A to F. 
 The galaxies spanned about two orders of magnitude in stellar masses from $\rm \approx 10^{9-11}~M_{\odot}$. 
All the galaxies were star-forming, with specific star formation rates (sSFRs) ranging between $\rm -10.2 < log(sSFR~yr^{-1})< -9.3$.  
 Their properties are listed in Table~1.  
 The SDSS-based group finders published by \citet{yang07} listed the halo mass of this group to be  $\rm 2.5\times 10^{12}~M_{\odot}$, although it is worth noting that this catalog did not list galaxies E and F as part of the group. These two galaxies, while having the right redshifts, were much farther away from the group center and may not have had the chance to virialize yet. Therefore, they were left out of the calculations for estimating the halo mass.

Our sight line toward QSO SDSS J151237.15$+$012846.0  pierced the heart of the group. It lay at an impact parameter of 65~kpc from the galaxy SDSS~J151243$+$012753 (galaxy A) - the brightest galaxy in the group. The closest galaxy to the sight line was galaxy B at an impact parameter of 64~kpc. Galaxy C was at an impact parameter of 111~kpc.  
 Although this QSO sight line was initially observed for the COS-GASS survey \citep{borthakur15a, borthakur16b} to probe the circumgalactic medium of galaxy A, it was left out of the sample, as the sight line also probed the halos of galaxies A, B, and C simultaneously. Because the three galaxies had the same redshift, it was not possible to ascertain the origin of the absorbers without additional data on the galaxies, which were not available to the COS-GASS survey.

\section{QSO Slight Line Probing a Low-$z$ Galaxy Group}

\subsection{HST Observational Setup and Data Analysis}

We obtained the ultraviolet spectrum of the QSO using the high-resolution grating G130M on the Cosmic Origins Spectrograph (COS) 
aboard the {\it Hubble Space Telescope (HST)}. The spectrum covered the rest frame wavelengths from 1157 to 1472$\rm \AA$ at a spectral resolution of 15$-$20~\kms.
All absorption lines, including those associated with intervening systems, were identified interactively.
Absorption features with equivalent widths $\rm W \ge 3 \sigma$, were identified as ``true features" and were further analyzed. 
The continuum was set using absorption-free regions of the spectrum within $\pm$1000~kms$^{-1}$ (in some cases up to $\pm$3000~kms$^{-1}$) of the absorption feature and subsequently fitted with a Legendre polynomial of order between 1 and 5, similar to the procedure used by \citet{sembach04}. This ensured that the continuum was described in the neighborhood of the absorption feature precisely and nullified any slow variations in the QSO's spectrum. This step is crucial for identifying weak features where small-scale continuum variations can be significant.  We also iterated continuum placement multiple times to reveal any weak 3$\sigma$  features. 

Absorbers within 600~km~s$^{-1}$ of the group redshift were identified as associated absorption of the group. They were then cross-referenced with other metal-line transitions of similar ionization states to confirm the velocity profile. All the absorption features in the QSO spectrum - including the intervening systems, QSO's intrinsic absorption, and the Milky Way's 
interstellar medium (ISM) - were identified to assess possible blending of the absorption features. To arrive at the measurements, we deblended the absorption features that had blending signatures by bootstrapping multiple optically thin transitions of the same species to arrive at the measurements.

\begin{figure}
\includegraphics[trim=5mm 0mm 0mm 0mm,  clip=true, scale=0.5]{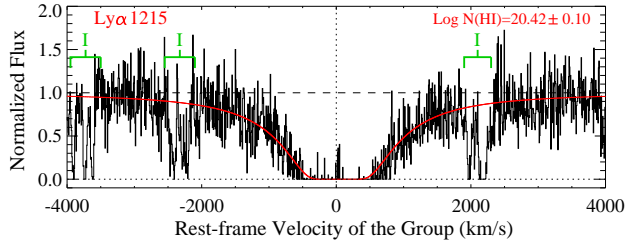}
\caption{COS spectrum showing the \Lya absorption features at z=0.029268. The zero-velocity marks the redshift of the group. The column density of log~N(HI)= 20.42$\pm$0.10 was estimated by fitting a Voigt profile to the absorption feature. The fit is shown in red. We detected a weak \Lya emission at the centroid of the \Lya profile at 44~\kms. The similarity in  velocity distribution of the \Lya absorption and emission indicates that both the absorption and the emission were produced in the same structure. Intervening absorbers are marked with `I' in green.}
 \label{COS_spectra_Lya}
\end{figure}

\begin{figure*}
\hspace{-1cm}

\includegraphics[trim=5mm 0mm 0mm 0mm,  clip=true, scale=0.45]{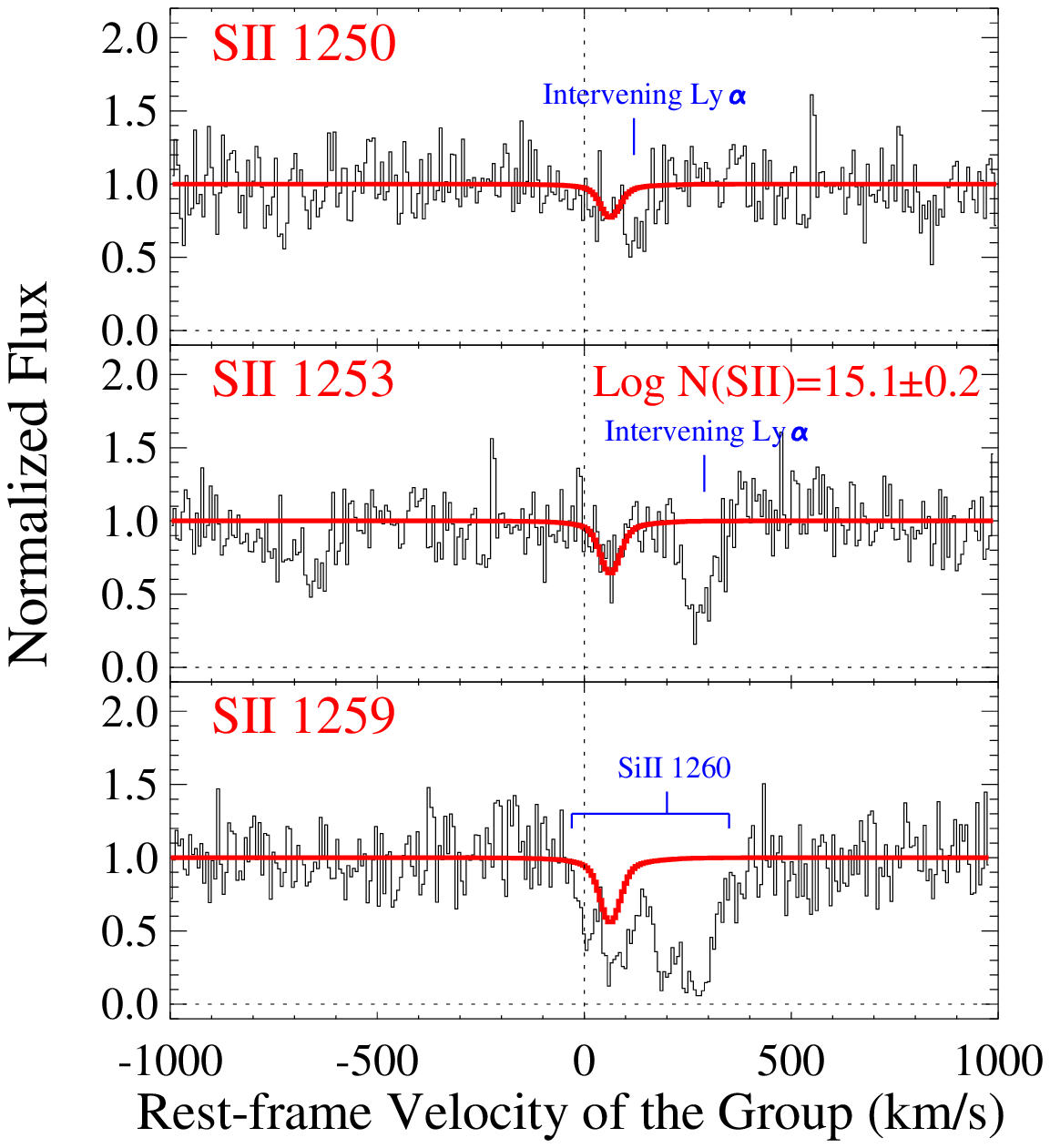}
\includegraphics[trim=5mm 0mm 0mm 0mm,  clip=true, scale=0.45]{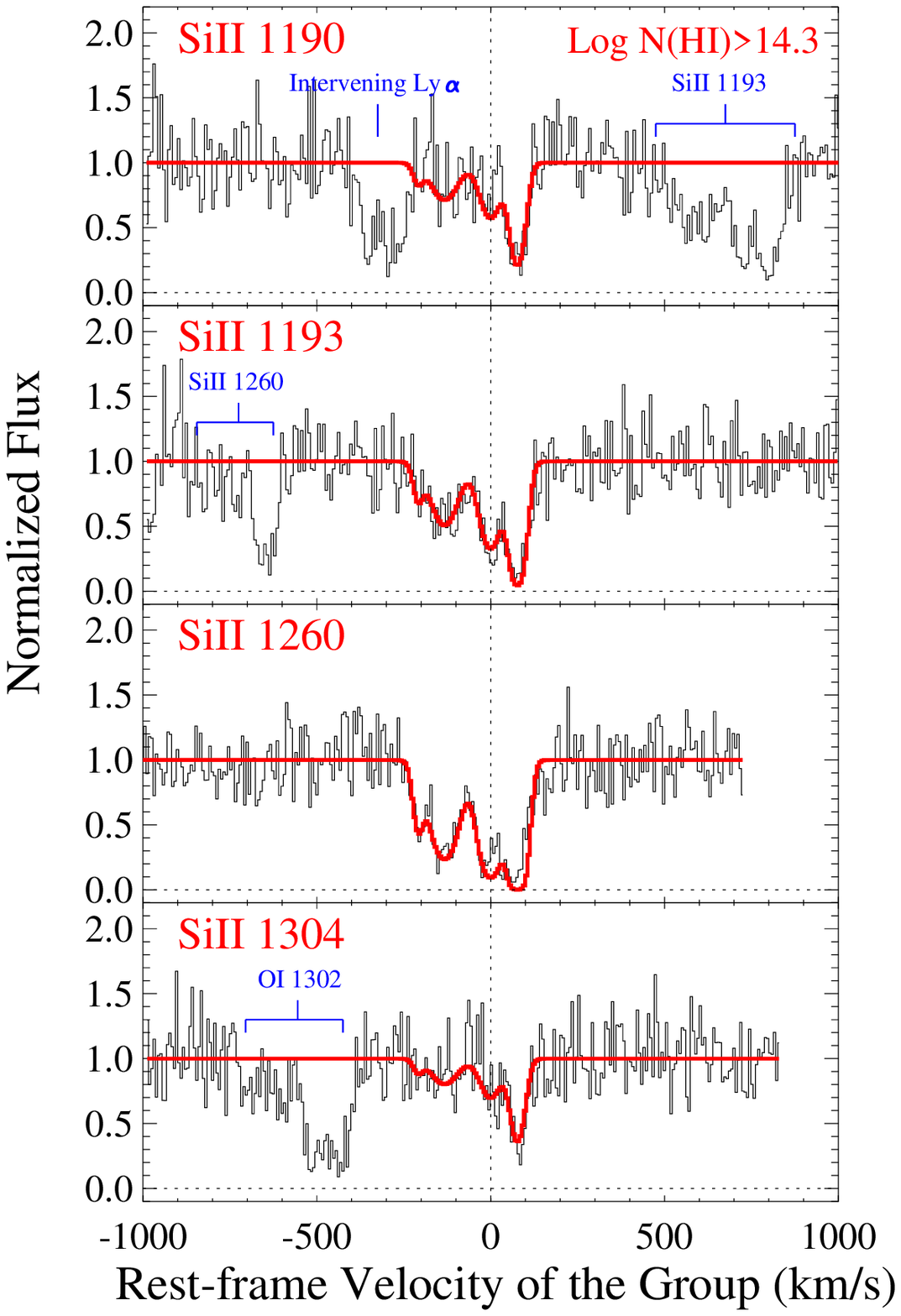}
\includegraphics[trim=5mm 0mm 0mm 0mm,  clip=true, scale=0.45]{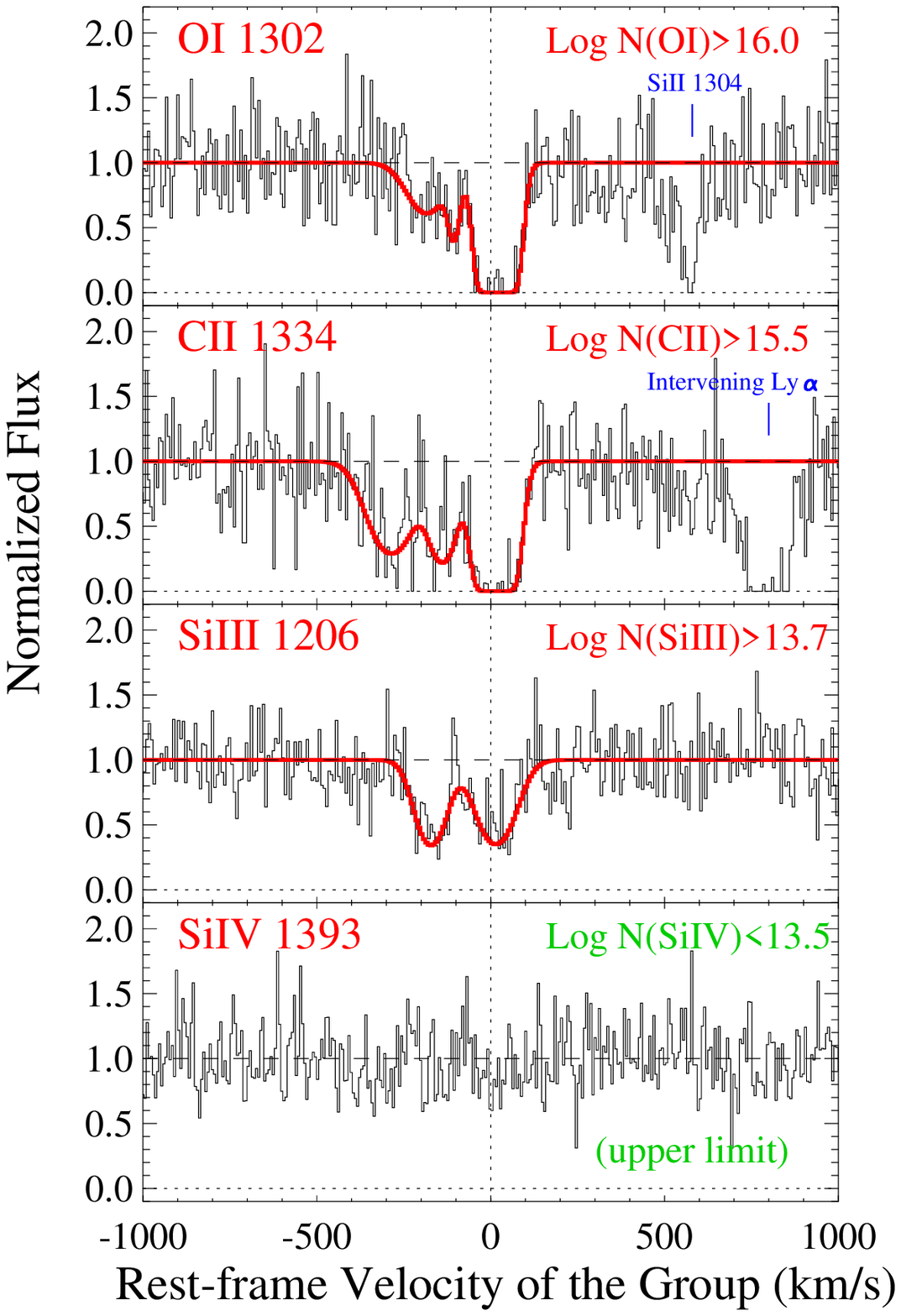}
\caption{COS spectra showing the metal-line transitions of the DLA system in absorption at a redshift of the group (z=0.029268). We detected a wide variety of transitions tracing low ionization gas \textsc{H~i}, \textsc{S~ii},  Si~\textsc{ii}, \textsc{O~i}, \textsc{C~ii}, and intermediate ionization traced by Si~\textsc{iii}. We did not detect Si~\textsc{iv} suggesting that the gas was primarily neutral. Left: the optically thin \textsc{S~ii} lines that allowed us to measure the column density of \textsc{S~ii} in the absorbing gas. The lines $\lambda$1250 and $\lambda$1259 were blended and the Voigt profile was obtained from fitting the transition $\lambda$1253. It was used to create models for the absorption profile from the other two transitions. The models are shown in red.  Middle: optically thin and thick components observed in the four transitions of Si~\textsc{ii}. Si~\textsc{ii} 1304 was optically thin. The other transitions were primarily optically thick. The  two blueward (negative) components were partially optically thin and showed the expected ratio of oscillator strengths. Right: Optically thick low ionization transitions such as \textsc{O~i} and \textsc{C~ii} and intermediate-ionization transition Si~\textsc{iii}. The higher ionization state as traced by Si~\textsc{iv} transition was also included in the panel. \textsc{Si~iv} was not detected in our data, and an upper limit was derived. }
 \label{COS_spectra_metals}
\end{figure*}

\subsection{Detection of a Damped Ly$\alpha$ System }

We detected \HI (\Lya) $\lambda$ 1216 at the redshift of the group. 
Figure~\ref{COS_spectra_Lya} presents the \Lya profile as a function of the rest frame velocity of the group ($v=0$ corresponds to $z=$0.0293). The \Lya  absorption feature exhibited prominent damping wings. The velocity spread of the feature was over 4000~\kms with the trough covering 700~\kms. A Voigt profile fit to the absorption feature gave a column density estimate of $\rm log~N(HI)=20.4$, thus making it a DLA system\footnote{The same value was also estimated by \citet{N16} in the archival study of DLA system for this DLA.}. The velocity centroid was found to be at an offset of 71~\kms from the group redshift.  
This is not significant, considering the fact that uncertainty in the redshift estimation for the group was of a similar order.

We also detected \Lya emission precisely at the center of the \Lya profile at 68~\kms i.e. $\rm z_{DLA}=0.291$. This indicates that the emission arose from the same structure as the absorption \citep[similar to the discussed by][]{wolfe05,rahmani10,rauch11}. The DLA acted as a coronagraph, blocking the light of the background QSO, thus enabling the detection of \Lya emission from fainter objects at $z = z_{DLA}$. The emission is most likely associated with the larger-scale \HI structure traced by the DLA. Unfortunately, the quality of our data are not good enough to pursue kinematic analysis of the emission profile. 

The measurements for the \Lya transition are presented in Table~2. The equivalent widths reported in the table were measured over all the observed components for each of the species.  
Using curve-fitting techniques similar to those used by \citet{tripp08,tumlinson13}, we fitted Voigt profiles to the absorption features and derived the velocity centroids and b-values from the fits using the software by \citet{fitzpatrick_spitzer97}. The line-spread function of the COS G130M grating was folded into the fitting procedure. The number of components that were fitted was determined by visual inspection. For species with multiple transitions, the fit was derived by fitting all the transitions optimally. The \HI column density and velocities were derived from \Lya alone, as we did not have access to any other transitions.
The uncertainties in the measurements were estimated using the error analysis method presented by \citet{sembach92} and  included continuum placement uncertainties along with Poisson noise fluctuations and flux zero-point uncertainties.

\subsubsection{Metal-line Transitions Associated with the DLA}

We detected  multiple metal-line species associated with the DLA including \ion{C}{2} $\lambda$1334; \ion{O}{1} $\lambda$1302; \ion{Si}{2} $\lambda\lambda$ 1190, 1193, 1260, and 1304; \ion{Si}{3} $\lambda$ 1206; and \ion{S}{2} $\lambda\lambda$ 1250, 1253, and 1259. 
Figure~\ref{COS_spectra_metals} presents the COS spectra of the metal-line transitions. 
Most of the absorption features associated with low ionization states, such as \ion{O}{1}, \ion{C}{2} , and  \ion{Si}{2}, had multiple components and were optically thick.

\ion{O}{1}~$\lambda$1302 showed three components at 27, -177, and -99 \kms (in the order of their strength) and had a total column density of $\rm log~O(I)>16.0$. The presence of strong \ion{O}{1} that has an ionization potential of 13.6~eV (very close to the ionization potential of \ion{H}{1}) suggests that the gas is primarily neutral. The offset in velocity centroids between the DLA and the strongest \ion{O}{1} component (at 68~\kms vs. 27~\kms in the rest frame of the group) is not surprising given that the \ion{O}{1} feature is saturated with a broad trough, which adds significant uncertainty to the estimation. The same is the case for \ion{C}{2} where we detected two components at 16 and -160~\kms. The strongest feature is saturated, thus giving us a lower limit of the column density of $\rm log~(CII)>15.4$.

\ion{O}{1}~$\lambda$1302 showed three components at 27, -177, and -99  \kms) (in the order of their strength) and had a total column density of $\rm log~O(I)>16.0$. The presence of strong \ion{O}{1} with an ionization potential of 13.6~eV, which was  very close to the ionization potential of H1, suggested that the gas was primarily neutral. The offset in velocity centroids between the DLA and the strongest O1 component (at 71~\kms vs. 27~\kms in the rest frame of the group) was not surprising, given that the \ion{O}{1} feature was saturated with a broad trough, adding significant uncertainty to the estimation. The same was the case for \ion{C}{2} where we detected two components at 16 and -160~\kms. The strongest feature was saturated, indicating a lower limit to the column density of $\rm log~(CII)>15.4$.

Our data covered three adjacent ionization states of silicon, namely \ion{Si}{2}, \ion{Si}{3}, and \ion{Si}{4}. 
In addition, we covered four transitions of \ion{Si}{2} - $\lambda\lambda$1190, 1193, 1260, and 1304, with different oscillator strengths.  \ion{Si}{2} $\lambda$1260 had the highest oscillator strength, while \ion{Si}{2} $\lambda$1304 had the lowest. This allowed us to evaluate the saturation level in the profiles.  
We detected four components in the \ion{Si}{2} profile at 70, -7, -139, and -125~\kms listed in order of their strengths. They showed different levels of saturation. 
The strongest component was saturated in all the four transitions. The weaker components were optically thin and showed the expected ratio of absorption strengths corresponding to their intrinsic oscillator strengths. Therefore, we estimated $\rm log~(Si II)>14.3$. 

We detected two components in the \ion{Si}{3} absorption profile at 21 and -165~\kms respectively. The level of saturation of these features could not be determined accurately, although the best-fit Voigt profile suggested that they have been mildly saturated. Unlike the low ionization species, the \ion{Si}{3} absorption feature showed only two components and of almost equal strengths. This could be due to the blending of the features. Qualitatively, the \ion{Si}{3} profile was similar to the \ion{Si}{2}, \ion{C}{2}, and \ion{O}{1} profiles with two distinct absorption complexes at $\approx$ 21 and -165~\kms. However, the ratio of the relative strengths of these two complexes (i.e., the ratio of their equivalent widths) increased from $\approx$ 0.4 to 1.0 for  \ion{O}{1} through \ion{Si}{3} with \ion{Si}{2} and \ion{C}{2} showing intermediate ratios.
Interestingly, this was the same order as the ionization potential of each of these species. Therefore, we conclude that we were seeing two absorbing complexes with slightly different ionization states. The gas associated with the blue component centered at -165~\kms is at a higher ionization state than that of the gas associated with the red component centered at 21~\kms. 

The two components need not be spatially connected. The strongest component was seen in all the low ionization transitions and is most probably associated with the DLA. The weaker and higher ionized component could be a  chance projection of a background or foreground cloud with respect to the DLA. We discuss the differences between the two components later. 

The highest ionization state of silicon in our wavelength coverage was \ion{Si}{4}. We did not detect any absorption at the expected wavelength of \ion{Si}{4}. The estimated upper limit for \ion{Si}{4} column density was $\rm log~SiIV< 13.5$. The nondetection of \ion{Si}{4} suggested that although the blue component seen in \ion{Si}{3} may be tracing gas at a higher ionization state than the red component, the overall ionization state of the gas in both the components was fairly low ($<<$ 33.5~eV needed to ionize \ion{Si}{3}). Furthermore, this observation was consistent with \ion{Si}{3}$\lambda$1206 being the strongest absorption line tracing metals in the warm circumgalactic medium of galaxies \citep[also see][]{collins09,  shull09, lehner12, lehner15,borthakur16b, richter16}. It was expected to trace even the smaller pockets of warm neutral to semi-ionized gas. 

In addition to the commonly detected metal-line transitions in absorption, the higher column density of the system does make it possible to detect transitions of other species with much weaker oscillator strengths. In this case, we detected optically thin transitions of low ionization species such as \ion{S}{2} $\lambda\lambda$1250, 1253, and 1259. The \ion{S}{2} $\lambda\lambda$1250, and 1259 transitions showed blending with intervening \Lya and \ion{Si}{2} $\lambda$1260, leaving the \ion{S}{2} $\lambda$ 1253 as our best candidate for column density measurements. 
The best-fit parameters obtained from the Voigt profile fit to this transition were used to generate expected profiles of t\ion{S}{2} $\lambda$ 1253 transitions, which matched the observed features perfectly.
 This confirmed that the \ion{S}{2} $\lambda$ 1253 transition was in the linear part of the curve of growth with a column density of $\rm log~(S II)=15.1$ and was not affected by saturation effects.

Our data also covered the observed wavelengths for \ion{N}{5}~$\lambda\lambda$1239, and 1243, but we did not detect any absorption. This result was consistent with the nondetection of \ion{Si}{4}. The data did not cover the observed wavelength for \ion{C}{4}  or \ion{O}{6} and hence no measurements could be made. It is worth noting that the nondetection of \ion{Si}{4} and \ion{N}{5} significantly reduces the likelihood of a \ion{C}{4} detection. The same can be extended to \ion{O}{6} as well, although the hot ($\rm 10^{5.5-6}~K$) intragroup medium expected in groups may contribute to the \ion{O}{6} profile \citep[][and references therein]{mulchaey96, stocke14, stocke17, faerman17}.
We list all the measurements for the transitions discussed above in Table~2, including equivalent widths, velocity centroids, b-values, and column densities.

\subsection{Metallicity and Ionization Correction}

Although we did see multiple absorption components in the metal-line transitions, it was not possible to identify associated neutral hydrogen column densities. This was the case for all DLAs where the \Lya absorption was broad and symmetric \footnote{Asymmetric profiles have been successfully deblended by \citet{prochaska09}},  as well as spread over a few thousand kilometers per second, thus washing out any signs of individual components. 
We assumed that the system could be considered as the sum of the components, and we evaluated an average metallicity and ionization state of the gas.

 \begin{figure}
 \includegraphics[trim=  20mm  120mm  0mm  0mm,  clip=true, scale=0.45]{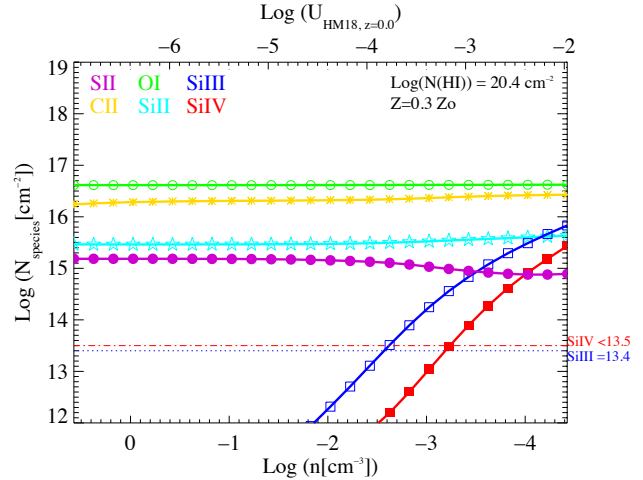}
 \caption{Photoionization modeling, using numerical code CLOUDY \citep{cloudy}, shows the column densities of various species for a slab of gas with neutral \textsc{H~i} column density of $\rm 2.5\times 10^{20}~cm^{-2}$. The column densities for the transitions were represented with different colors as a function of density and ionization parameter ($U_{HM18,z=0.0}$). The ionizing source was assumed to be the cosmic ultraviolet background as described by \citet{haardt_madau18}. The observed column density of Si~\textsc{iii} (component at 22~\kms) and the limiting column density of Si~\textsc{iv} indicated that the density of the cloud was $\rm 0.0025-0.00063~particles~cm^{-3}$. As expected, the modeled column densities of most low ionization species were an order of magnitude higher than the limits. The exception was \textsc{S~ii}, which was an optically thin transition and showed little variation as a function of ionization parameter. }
 \label{cloudy}
\end{figure}

The absence of absorption associated with higher ionization transitions such as \ion{Si}{4} and \ion{N}{5} suggested that the gas was in a low ionization state. This was corroborated by optically thick transitions of \ion{C}{2} and \ion{Si}{2}.
One of the most suitable elements for identifying the metal content is the element sulfur as it does not deplete onto dust and lies on the $\alpha$-element synthesis sequence \citep{rafelski12}.
Therefore, we used optically thin \ion{S}{2} to estimate metallicity of the DLA, as most of the sulfur was expected to be in this ionization state. This yielded a metallicity estimate of $\rm 0.33^{+0.17}_{-0.11}~Z_{\odot}$. The observed column densities of silicon and carbon are 13 and 8 times lower than expected. This can be attributed to the uncertainty of the column density  estimates of silicon and carbon owing to the saturation of the absorption lines and our assumption that the detected transitions were the dominant ones, as well as due to depletion. 

We ran the radiative transfer code CLOUDY \citep{cloudy}, to generate ionization models to estimate the density of the cloud. The models were generated assuming that the cloud is a static slab of gas that is being irradiated by the cosmic ultraviolet background from all sides. These simulations used the latest corrections to the UV background, i.e., ionization parameter,  U=$\rm \Phi_{HM}/(c~n_H)$, where $\rm \Phi_{HM}= 6\times 10^3~photons~cm~^{-2}~s^{-1}$  \citep{haardt_madau18}.
The models were run for a wide range of gas densities, $\rm n=0.3- 3\times 10^{-5} ~cm^{-3}$, and ionization parameters $\rm  log(U)\approx -2~to~-7$. The neutral hydrogen column density was fixed to log(HI)=20.4, as measured by fitting the Voigt profile to the \Lya absorption feature. 


The density of the cloud came  from the column densities of the two adjacent ionization states of silicon - \ion{Si}{3} and \ion{Si}{4}. While \ion{Si}{3} $\lambda$1206 had a column density of $\rm log~N(Si~III) \ge 13.4$ (the main component), no absorption was seen for \ion{Si}{4} thus resulting in an upper limit of $\rm log~N(Si~IV)\le 13.5$. These levels are marked in Figure~\ref{cloudy} with the blue (for \ion{Si}{3}) and red (for \ion{Si}{4}) dashed lines. This limited the average densities for the gas cloud to be less than $\rm 2.5\times10^{-3} ~particles~cm^{-3}$. This implied that the DLA was at least 29~kpc along the line of sight.

It is worth noting that most of the low ionization transitions, such as \ion{O}{1}, \ion{Si}{2}, and \ion{C}{2}, are modeled to have significantly larger (1-0.3 dex higher) column densities than what was measured. However, that was expected as these transitions were in the optically thick regime as evident by their saturated absorption profiles, and therefore the measured values are the lower limits. In general, the measured column densities (and limits) were consistent with the gas in the DLA being photoionized by the cosmic ultraviolet background.

 \begin{figure*}
 \hspace{1cm}
\includegraphics[trim=40mm 60mm 40mm 30mm,  clip=true, scale=0.90, angle=-90]{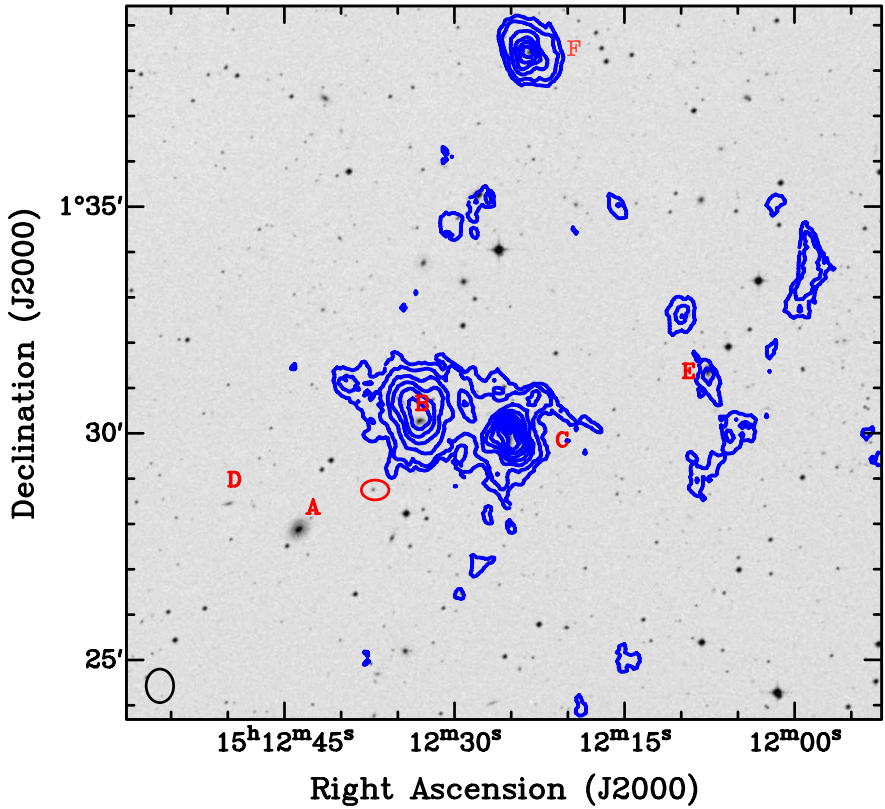}\\
 \caption{Digital Sky Survey (DSS-2) red-band image in gray scale overlaid with contours showing the VLA C$+$D configuration image tracing 21~cm \textsc{Hi} distribution associated with the group. The synthesized beam of 41.56$^{\prime\prime}\times$35.37$^{\prime\prime}$ is shown in the lower left corner. The contours represent the column densities of 5.3, 10.5, 21.0, 31.5, 42.1, 52.6, 63.8, and 74.4~$\times~\rm 10^{19}~cm^{-2}$. The background QSO is marked with a red oval.}
 \label{HI_image_2}
\end{figure*}

 \begin{figure*}
\includegraphics[trim=40mm 10mm 70mm 10mm,  clip=true, scale=0.8, angle=-90]{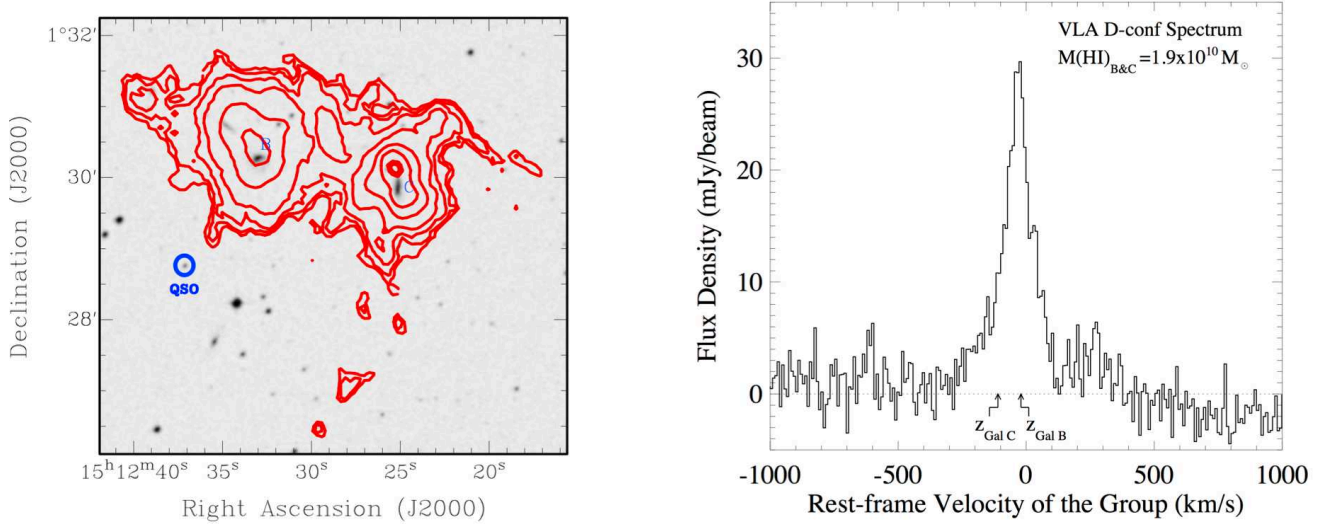}   
 \caption{Left: A zoom-in of Figure~\ref{HI_image_2} showing the Digital Sky Survey (DSS-2) red-band image in gray scale overlaid with contours showing the VLA C$+$D configuration image tracing 21~cm \textsc{Hi} distribution associated with galaxies B and C. The position of the QSO sight line is marked with a blue circle and labeled. The synthesized beam of the image is  41.56$^{\prime\prime}\times$35.37$^{\prime\prime}$. The contours represent column densities of 5.6, 7.5, 11.3, 18.8, 37.6, 56.3, and 75.1~$\times~\rm 10^{19}~cm^{-2}$.  Right: The VLA D configuration \textsc{Hi} spectrum of the complex. The D configuration map was the most sensitive (although of lower spatial resolution) and provided us with the best estimate of the \textsc{Hi} mass. The \textsc{Hi} mass of the complex was measured to be $\rm 1.9 \pm 0.1 \times10^{10}~M_{\odot}$.}
 \label{HI_image_BC}
\end{figure*}

 \begin{figure*}
\includegraphics[trim=40mm 60mm 40mm 65mm,  clip=true, scale=0.57, angle=90]{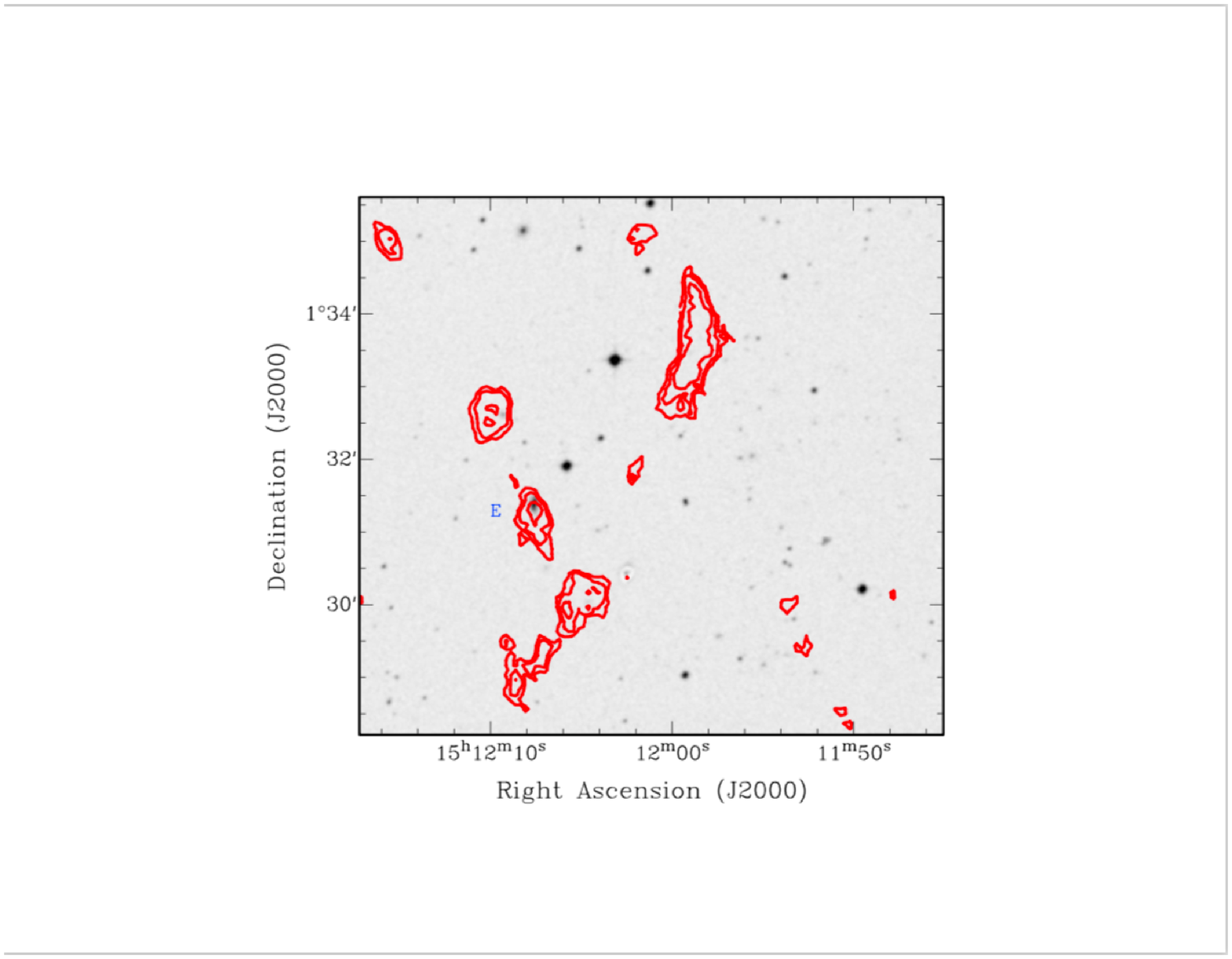} 
\includegraphics[trim=0mm 0mm 0mm 0mm,  clip=true, scale=0.43]{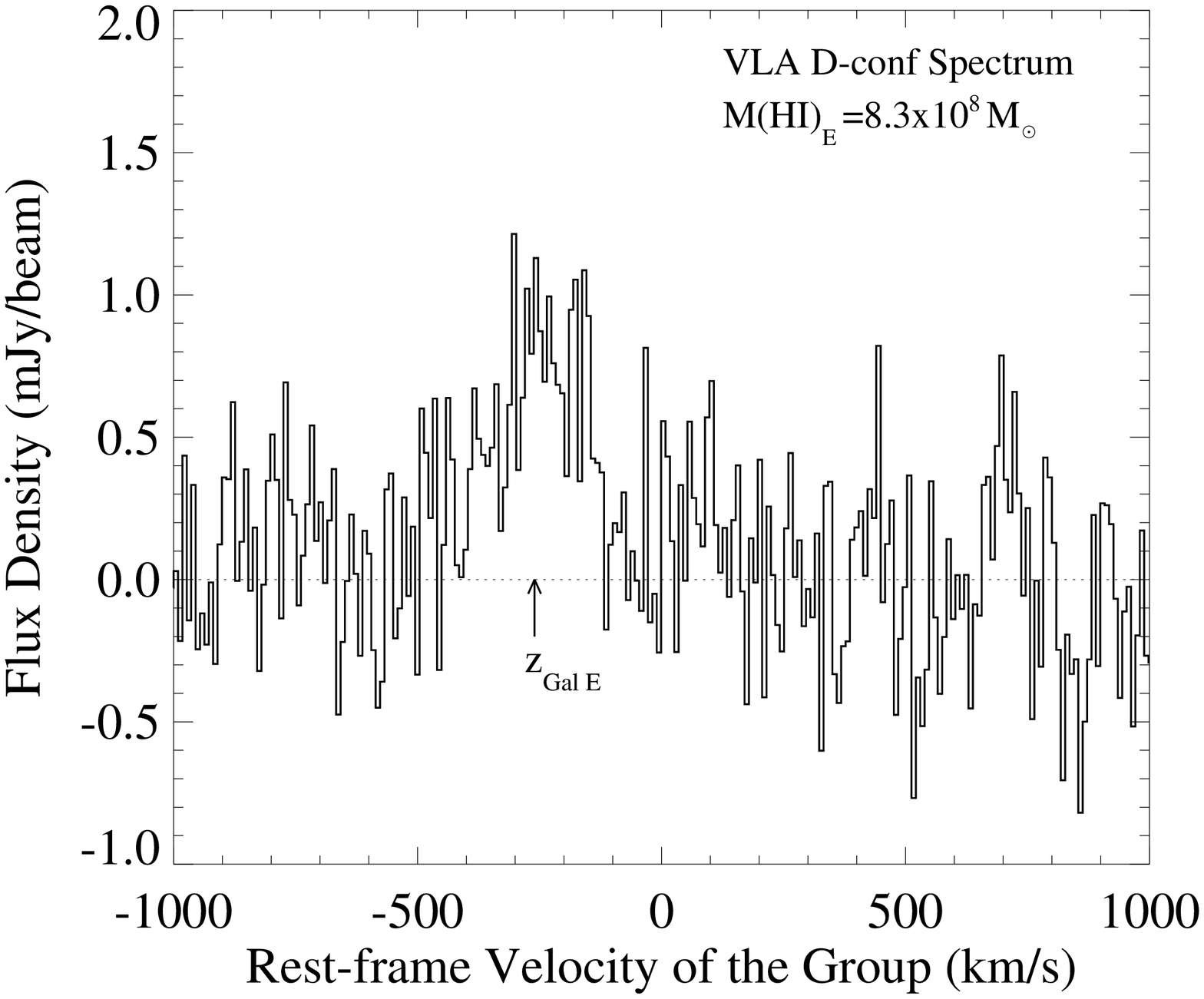}
 \caption{Right: A zoom-in of Figure~\ref{HI_image_2} showing the  DSS-2 red-band image in gray scale overlaid with contours showing the VLA C$+$D configuration image tracing 21~cm \textsc{Hi} distribution associated with galaxy E. The contours represent \HI column densities of 5.6, 7.5, 11.3, 18.8, 37.6, 56.3, and 75.1~$\times~\rm 10^{19}~cm^{-2}$.  Right: The \textsc{Hi} spectrum of the  galaxy E showing a total \textsc{Hi} mass of $\rm 8.3 \pm 0.8  \times 10^{8}~M_{\odot}$. A first-order polynomial was fitted to remove a dipping continuum at the edges. The order of the polynomial fit did not change the \textsc{Hi} mass by more than a couple of percent. }
 \label{HI_image_E}
\end{figure*}

 \begin{figure*}
\includegraphics[trim=40mm 60mm 40mm 60mm,  clip=true, scale=0.55, angle=90]{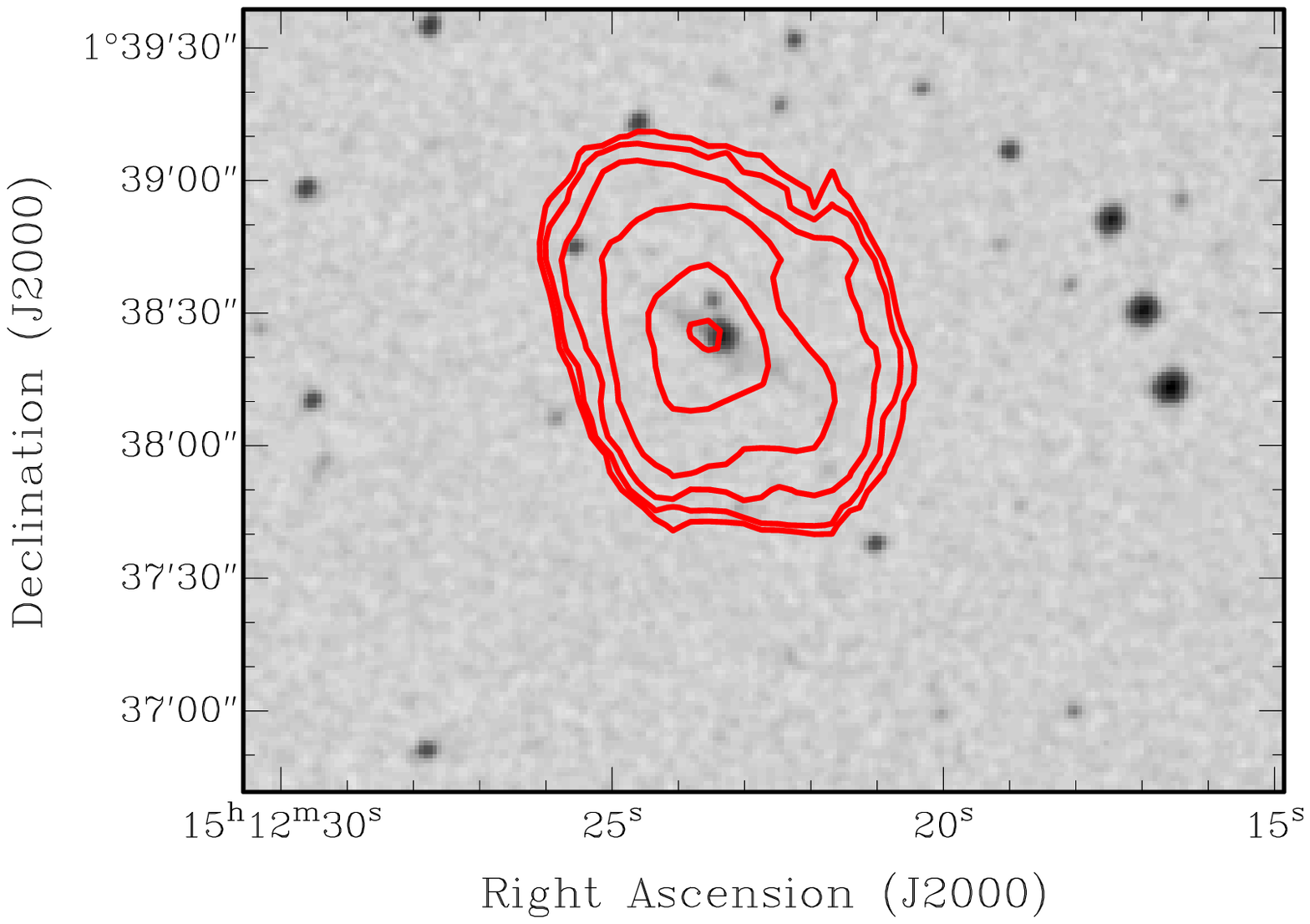} 
\includegraphics[trim=0mm 0mm 0mm 0mm,  clip=true, scale=0.4]{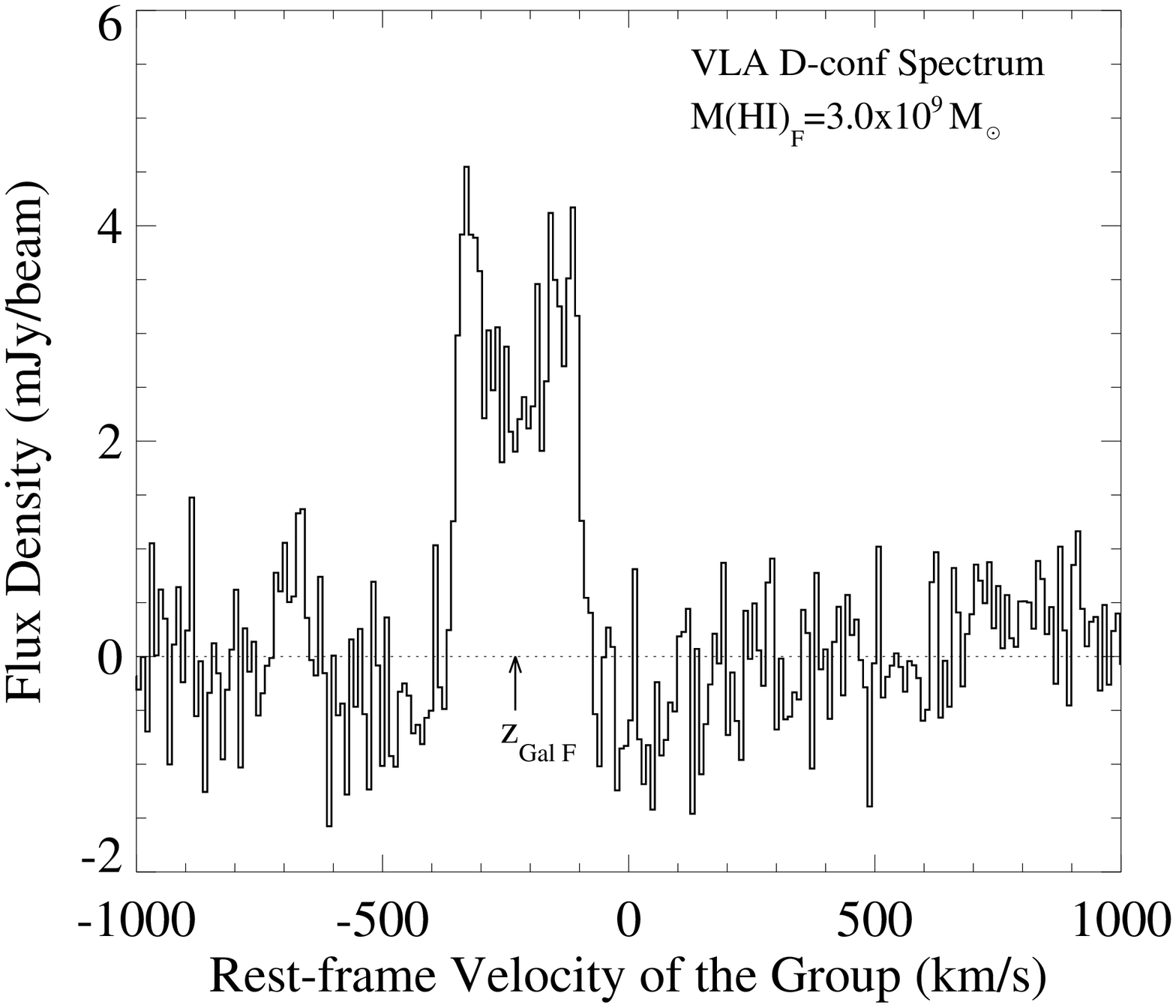}
 \caption{Left: Digital Sky Survey (DSS-2) red-band image in gray scale overlaid with contours showing the VLA C$+$D configuration image tracing 21~cm \textsc{Hi} distribution associated with galaxy F. The contours represent the column densities  5.6, 7.5, 11.3, 18.8, 37.6, and 56.3~$\times~\rm 10^{19}~cm^{-2}$.  Right: The VLA D configuration spectrum of galaxy F showing a double-horn \textsc{Hi} profile indicative of a rotating disk. The measured \textsc{Hi} mass of the galaxy was  $3.1 \pm 0.07 \times 10^{9}~\rm M_{\odot}$. The \textsc{Hi} dynamics did not show any sign of perturbation, suggesting that the galaxy was a new addition to the group. It might have just entered the potential and not yet had its first pass through the group center. }
 \label{HI_image_F}
\end{figure*}

 \section{Atomic Gas Associated with the Group}
 
 The redshift of this system allows us to map the neutral \HI  in its environment.  Since the discovery of this DLA is unbiased in terms of its neighborhood, the insights from studying its environment could shed light on higher-redshift systems that cannot be imaged in \HI.
 Therefore, we conducted single-dish and interferometric observations of this system in \HI 21~cm spin transition to spatially connect the DLA to the atomic gas structures in its neighborhood.

\subsection{Observations}
 
One of the member galaxies of this group, galaxy A, was observed with the Arecibo telescope as part of the GALEX Arecibo SDSS Survey \citep[GASS;][]{catinella13}. This is the most massive galaxy in this group and lies at an impact parameter of 65~kpc from the sight line. The \HI mass detected in the Arecibo spectra  (beam size of $\rm 3.5^\prime$) centered at the position of galaxy~A  was $\rm 1.9\times 10^{9}~M_{\odot}$ (GASS~ID~7813) . However, the Arecibo beam was much larger than the projected size of the galaxy and covered a significant part of the group. Therefore, to characterize the \HI distribution more precisely, we conducted observations with the NSF's Karl G. Jansky Very Large Array (VLA\footnote{The National Radio Astronomy Observatory is a facility of the National Science Foundation operated under cooperative agreement by Associated Universities, Inc.}). In the rest of the paper, we will focus on the VLA data. We also obtained \HI spectra with the Green Bank Telescope (GBT), which is discussed in the Appendix.

We conducted VLA D  and C configuration \HI 21~cm imaging of this group under programs 14A-561 and 14B-057 for 4 and 2 hr, respectively. The data were obtained in both configurations in order to gain both sensitivity (from the D configuration) and resolution (from the C configuration). 
The data were obtained in a dual-polarization mode with the WIDAR correlator. A spectral resolution of 2.3$\rm ~km~s^{-1}$ was achieved. The data were analyzed with the Common Astronomy Software Applications (CASA) package of the National Radio Astronomy Observatory (NRAO). 
The data from each configuration were calibrated individually and then combined for higher resolution and high signal-to-noise ratio. The combined data (C$+$D hereafter) yielded an rms noise of 2.36$\times \rm 10^{-2}$~Jy/beam~$\rm \times ~km~s^{-1}$ corresponding to a column density of 1.8$\rm \times10^{19}~cm^{-2}$. 
The synthesized beam size of the combined data was 41.56$^{\prime\prime}\times$35.37$^{\prime\prime}$ at a position angle of 5.72$^\circ$. 

The VLA C$+$D \HI image is presented in Figure~\ref{HI_image_2} as blue contours overlaid on the optical image of the group. The member galaxies are identified, and the background QSO is marked with an ellipse. As the background QSO was not radio-bright, we did not suffer from any issues related to dynamic range.

\subsection{ Results from \HI imaging with the Very Large Array}

We detected a total \HI mass of $\rm 2.5 \pm 0.1 \times 10^{10}~M_{\odot}$ in this group. We detected \HI associated with galaxies B, C, E, and F. We did not detect \HI emission from galaxies A and D. 

Most of the \HI mass of the group was in a massive \HI cloud of $\rm 1.9 \pm 0.1 \times 10^{10}~M_{\odot}$ enveloping galaxies B and C (Figure~\ref{HI_image_BC}). The \HI distribution peaked at the position of the two galaxies with a broad bridge connecting them. Tail-like extensions (at 3$\sigma$) were seen in the southeast direction. In addition, multiple finger-like structures were also detected at the edge of the \HI distribution. The QSO sight line lay close to one such extension; however, at the resolution and sensitivity of the data, we were unable to detect any \HI that covered the QSO sight line.

Galaxy E, with an \HI mass of $\rm 8.3\pm 0.8 \times 10^{8}~M_{\odot}$ was also surrounded by multiple elongated \HI structures. These could potentially be surviving remains of a tidal stream (Figure~\ref{HI_image_E}). The \HI in galaxy F was detected in a symmetric rotating disk centered at the optical galaxy (Figure~8). This was indicated by the double-horn spectral profile with an \HI mass of $\rm 3.0 \pm 0.07 \times 10^{9}~M_{\odot}$. The \HI in galaxy F was detected in a symmetric double-horn spectral profile with $\rm M(HI)=3.0 \pm 0.07 \times 10^{9}~M_{\odot}$, thus indicating the presence of a rotating disk centered at the optical galaxy.


 \begin{figure*}[t]
 \hspace{-0.5cm}
\includegraphics[trim=65mm 20mm 65mm 20mm,  clip=true, scale=0.77, angle=90]{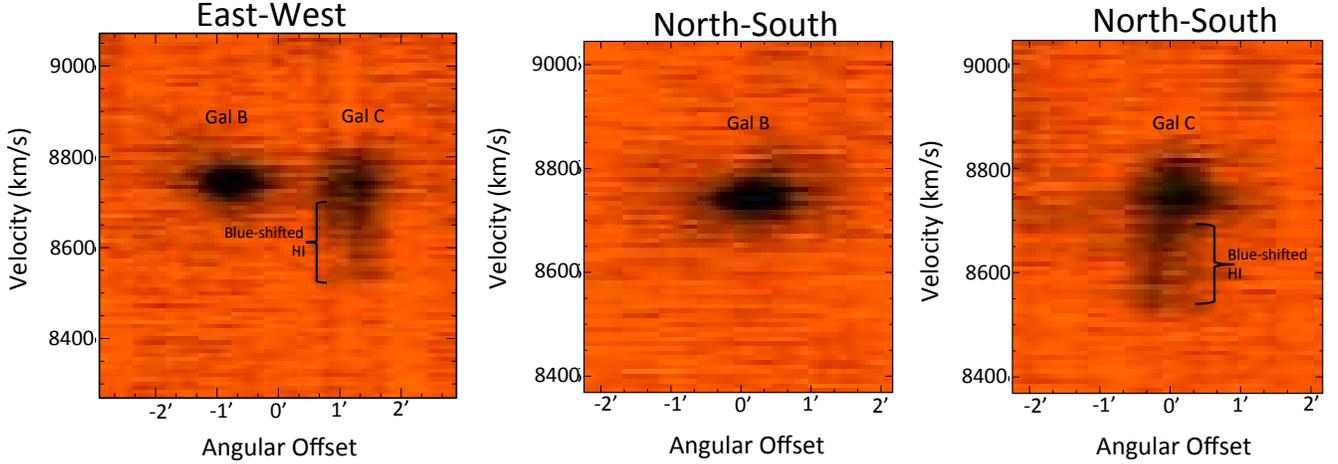}
\caption{Position-velocity plots show the \textsc{Hi}  distribution as a function of velocity through three position slices. The ordinate represents the observed velocity in the units of $\rm km~s^{-1}$. The left panel is a position cut from east-to-west passing through the centers of galaxies B and C. The middle and the right panels are north-to-south cuts for galaxy B on the left and for galaxy C on the right. While most of the \textsc{Hi} lay within 100~\kms of the peak at 8750~\kms (z=0.029), galaxy C showed an extended \textsc{Hi} emission tail reaching up to $\sim$200\kms blueward of the systemic.  }
 \label{HI_pos_vel_BC}
\end{figure*}

\subsubsection{The \HI Envelope around Galaxies B and C}

Here we discuss the kinematics of the \HI envelope around galaxies B and C. 
Figure~\ref{HI_pos_vel_BC} shows the position-velocity diagrams. The left panel is an east-to-west slice across the entire cloud passing through the centers of both the galaxies. The middle and the right panels show the north-to-south slices for galaxies B and C, respectively. The angular position (abscissa) in the plots was chosen arbitrarily. 
The strength of the \HI distribution peaked at the position of the two galaxies. The two peaks lie at the same velocity. \HI in galaxy C had a high-velocity component blueward of the systemic containing $\rm M(HI)_{blue}=2.2 \times 10^{9}~M_{\odot}$. The blueshifted tail was detected in all the slices that pass through galaxy~C, but was spatially unresolved. 
We discuss the nature of the blueshifted emission (inflow vs. outflow) in the next section.

\begin{figure*}
 \hspace{0.3cm}
\includegraphics[trim=0mm 00mm 0mm 0mm,  clip=true, scale=0.406, angle=0]{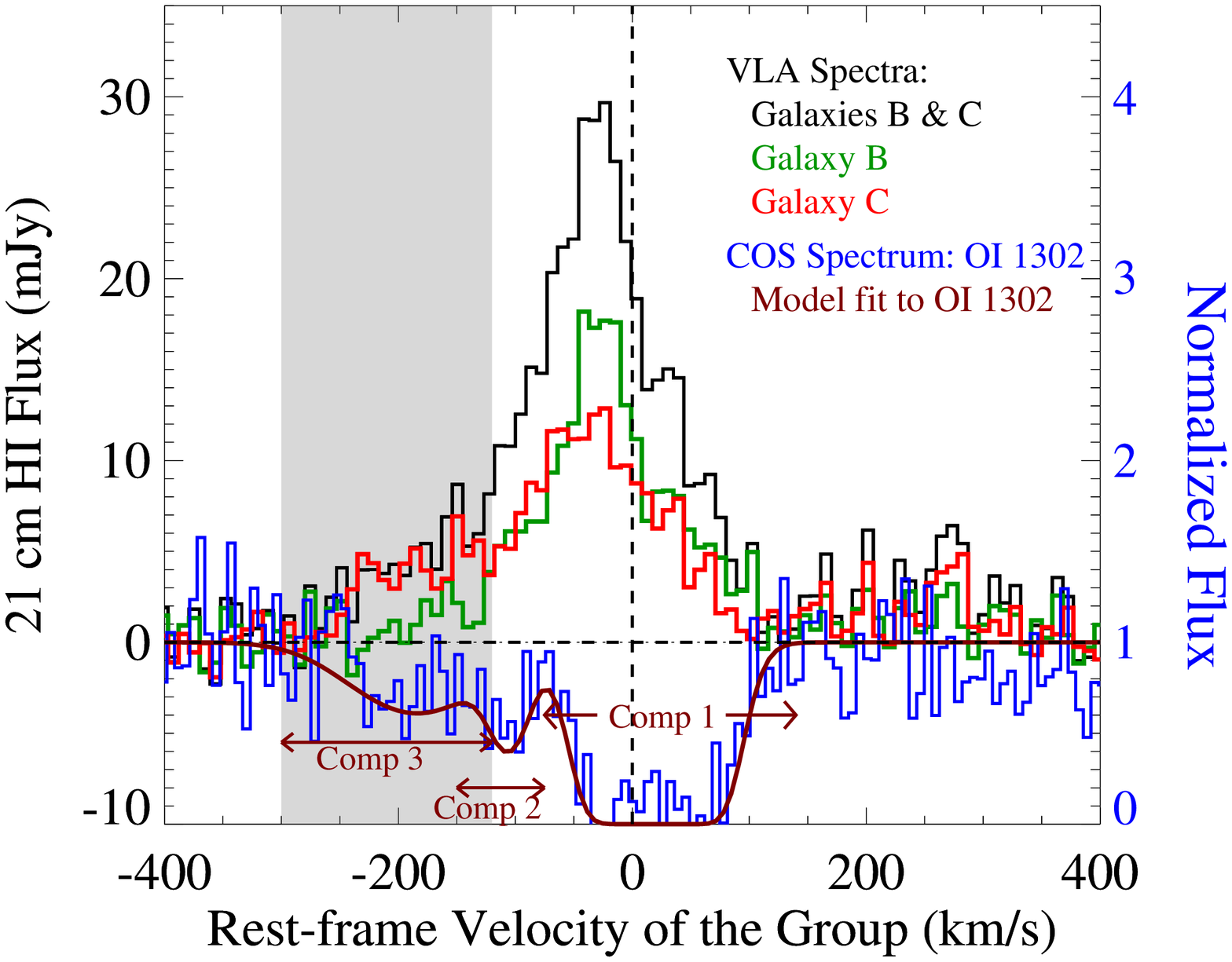} 
\hspace{0.5cm}
\includegraphics[trim=0mm 00mm 0mm 0mm,  clip=true, scale=0.405, angle=0]{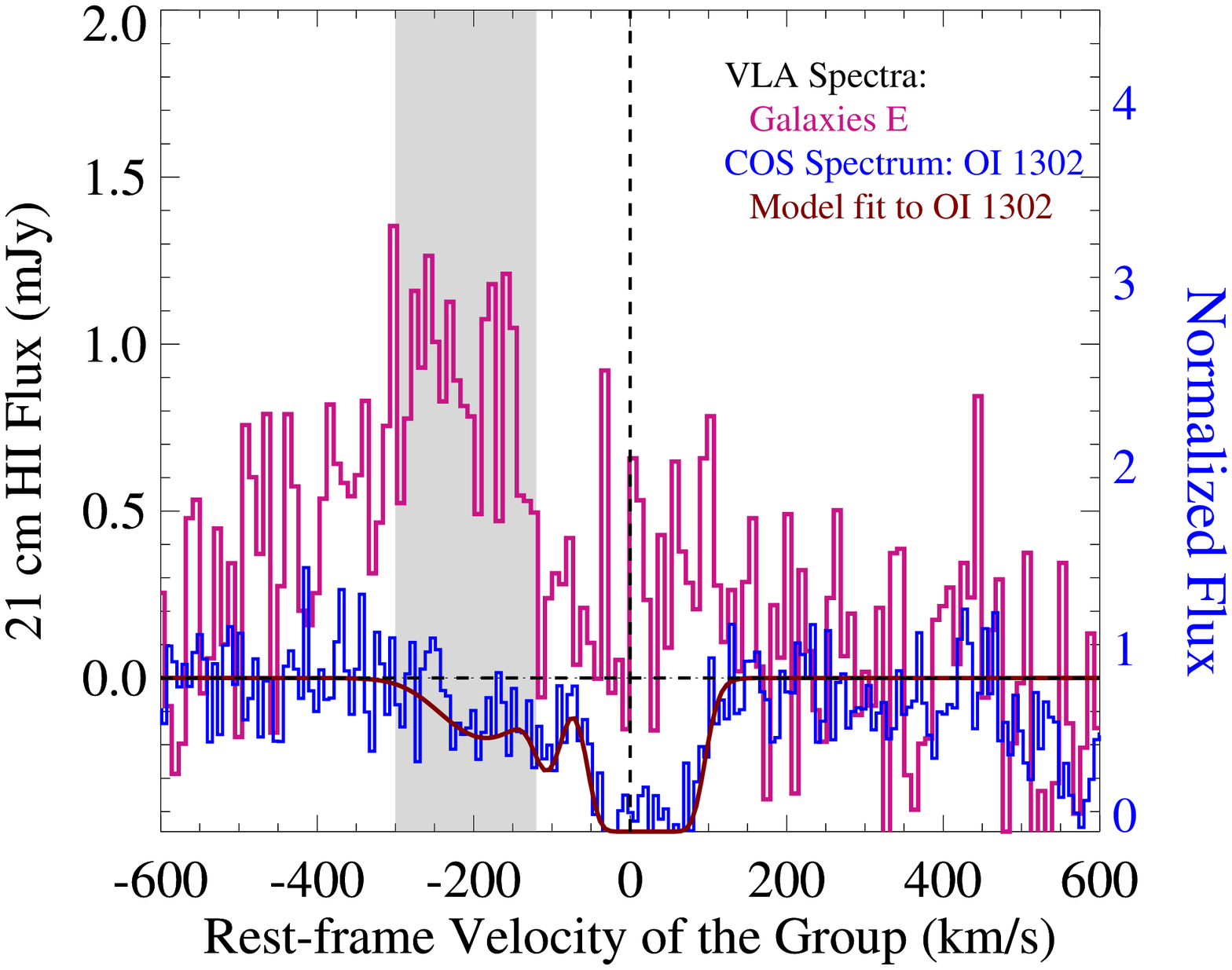} 
 \caption{LEFT: comparison of the kinematics of the \textsc{Hi} 21~cm line emission associated with galaxies~B$+$C, galaxy~B and galaxy~C with the DLA absorption profile traced by \textsc{Oi} $\lambda$ 1302. The abscissa refers to the rest frame velocity of the group. The ordinate on the left-hand side of the plot represents the \textsc{Hi}  21cm flux. On the right, the ordinate represents the normalized flux corresponding to the absorption profile shown in blue. The absorption profile was fitted with a Voigt profile with three components, which was shown in maroon. The kinematics of the blueward component matched  the blueshifted tail of \textsc{Hi} 21~cm emission seen in galaxy C. The region of interest is marked in gray. RIGHT: A comparison of the kinematics of the  \textsc{Hi} 21~cm line emission from galaxy~E and the bluemost component of the \textsc{Oi} $\lambda$ 1302 transition of the DLA. The wing of the blue component of the DLA extended to -300~\kms and overlapped with the peak of the \textsc{H~i} profile of galaxy~E.}
 \label{COS_VLA_comparison}
\end{figure*}

 In Figure~\ref{COS_VLA_comparison}, we show the comparison of the \HI 21~cm spectra with \ion{O}{1} absorption line tracing the kinematics of the DLA. The damping wings of the \Lya washed away the kinematic signatures of the gas in the DLA; consequently, we choose metal lines to trace the kinematics. We found \ion{O}{1} to be the best proxy for hydrogen, as it has an ionization potential of 13.62~eV, close to that of hydrogen. We compared the \HI spectrum of galaxies B$+$C (black), galaxy B (green), and galaxy C (red) to the absorption profile. The spectra of galaxies B and C were extracted from non-overlapping regions around the galaxies, whereas B$+$C was the total \HI of the complex.

Interestingly, the kinematic structure of the blueshifted emission was similar to the blueshifted wing of the \ion{O}{1} profile. The kinematic similarity suggested that the blueshifted tail of the \HI distribution of galaxy~C extended much farther out spatially. 
This is a surprising find because the QSO sight line was 112~kpc away from galaxy~C. 
In addition, the nature and location of the DLA were consistent with it being associated with the \HI envelope of galaxy~B. Therefore, the blueshifted gas in the QSO spectrum was a chance projection of a second cloud that was physically distinct from the DLA.  	
The filamentary \HI  structures around galaxy~E were also at similar velocities to the blueshifted emission. 
The extended nature of the blueshifted emission suggested that it traced a ``group-scale" structure.

\section{Discussion and Implications \label{sec:implications}}

\subsection{\HI Distribution in the Group }

The VLA imaging showed that most of the \HI associated with the group was contained in a large envelope around galaxies B and C. Even if each of them owned half of the \HI i.e. $\rm 0.95~\times~10^{10}~M_{\odot}$; that would be put them at the extreme end of the \ion{H}{1} mass function \citep{zwaan05, martin10}. Less than 20\% of all isolated galaxies have masses in this range or higher. Therefore, to have two of them in a group is fairly rare given that galaxies in groups are more likely to be \HI poor than galaxies in the field \citep{williams87}. 

Another interesting aspect of \HI distribution in this group is that no \HI emission was detected around galaxies A and D in our VLA imaging\footnote{The Arecibo spectrum detected an \HI mass of $\rm 1.9\times 10^9~M_{\odot}$ around galaxy~A. The gas could exist as faint diffuse gas that was not detected by the VLA, or the Arecibo beam might have picked flux from the B$+$C complex} despite being blue spirals. \HI-deficient galaxies are common in galaxy clusters and dense groups \citep[][and references therein]{yoon15,crone16}. In compact groups, \HI deficiency was related to the evolutionary stage, with highly evolved groups missing large fractions of their \HI \citep{VM05}. 
Expanding the theory to this group would imply that galaxies A and D had spent a significant amount of their lifetime in the group environment, where they were transformed into gas-poor galaxies. The most likely mechanisms for this transformation were the loss of their \HI via tidal interactions and ram pressure stripping \citep[and references therein]{rasmussen08, catinella17}.

\subsection{Possible Origins of the DLA}

We did not detect any \HI at the position of the DLA. The \HI emission column density (Figure~\ref{HI_image_2}) at the location of the DLA  was $\rm < 5 x 10^{19}~cm^{-2}$, while the fit to the \Lya profile gave a column density of $\rm 2.5 x 10^{20}~cm^{-2}$. The absorption strength was five times larger than the average column density measured in \HI emission. This is most likely an indication of beam dilution. Therefore, on scales of 20~kpc, the \HI distribution must have been highly patchy.

The location of the DLA with respect to the \HI in the group, the low metallicity and the low ionization state of the gas  in the DLA suggested a few possible avenues for its origin. The 21cm \HI maps indicated that the \HI distribution nearest to the QSO sight line was associated with galaxies B and C. Therefore, the DLA was probably associated with the extended \HI envelope of galaxies B and C. The kinematics of the DLA was also consistent with the kinematics of \HI in the B$+$C complex.

\subsubsection{Interstellar Medium, Tidal Debris, Dwarf Galaxy, or Cooling Flows?}

The closest \HI structure to our DLA is the \HI cloud enveloping galaxies~ B and C. 
The presence of finger-like structures fairly close to the QSO position advocated for a connection between the DLA and the ISM. The metallicity of the DLA was consistent with gas in the outer disks of galaxies \citep[][and references therein]{moran12}. In addition, the low ionization state of the gas in the DLA advocated for similar origin to the ISM.
 
Galaxy groups, being the sites of high galaxy densities with low velocity dispersion, are ideal for intense tidal interactions.
As a result, tidal structures are common in group environments.  These structures disperse the ISM of the individual galaxies into the intragroup medium. Could our DLA be associated with tidal debris? The finger-like structures of the ISM in galaxies B and C were most likely created via some tidal disturbance. Such morphologies are highly unusual in an undisturbed disk but are prevalent in slightly perturbed systems. We did not detect any tidal tails and bridges between galaxies B and C. Based on the close proximity of the two galaxies, we concluded that we were witnessing the early phase of galaxy interaction in this system. Therefore, it is likely that the DLA could be a tidally disturbed ISM cloud that was either attached to the \HI structure or free-floating in the intragroup medium. Such structures are likely to exist since the survival timescale for neutral gas in the intragroup medium at the virial temperature of $\rm 10^6~K$ is about 1 Gyr\citep{borthakur10a}. 

Another possibility satisfying the observed low metallicity and low ionization state of the gas was that the QSO sight line pierces through a dwarf galaxy. From our most sensitive D configuration \HI map, we estimate that the limiting \HI mass (corresponding to 3$\sigma$) of such a galaxy would have to be no more than $\rm 5 \times 10^{7}~M_{\odot}$ for an \HI line width of 100~\kms. 
The mass-metallicity relationship of dwarf galaxies predicted the stellar mass as log(M$_{\star})  \approx $ 8.8 \citep{berg12}. 
That would imply a low gas fraction $<10\%$ for a dwarf galaxy \citep[see][]{zhang09}. 

The presence of a foreground dwarf galaxy along the QSO sight line could be identified by identifying its emission lines that were superimposed on the QSO spectrum \citep[see]{york06, lundgren09, borthakur11}. In the past, H$\alpha$ had been used extensively for that purpose. We did not observe H$\alpha$ emission in the QSO spectrum at the redshift of the DLA, thus confirming that the dwarf galaxy (if any) is not star-forming. We estimated a star formation rate $\rm SFR < 0.0013~M_{\odot}~yr^{-1}$ within the SDSS fiber assuming a line width of 3.3~$\rm \AA$. This translates to an SFR surface density of  $\rm < 1.3 \times 10^{-4}~M_{\odot}~yr^{-1}~kpc^{-2}$ for the DLA hosts. This is very low for a normal star-forming dwarf galaxy \citep{borthakur11}. Although this hypothesis cannot be rejected with certainty, the likelihood of the DLA being part of a dwarf galaxy is quite low.

It is also unlikely that the DLA was produced by cooling flows \citep[][and references therein]{fabian94}. Those are commonly seen in galaxy clusters and X-ray-bright groups. 
This group was not X-ray bright, and hence cooling flows were unlikely. On the other hand, it is worth noting that the DLA had the same metallicity as that expected for the intragroup medium, i.e. $\approx \rm 0.3~Z_{\odot}$ \citep{peng14}. This is consistent with the idea of tidal structures feeding the intragroup medium.

\subsection{\Lya Emission and Its Possible Origin}

We detected weak \Lya emission at the trough of the DLA absorption profile. The flux in the emission profile was 4.9$\times \rm 10^{-17}~erg~s^{-1}~cm^{-2}$, corresponding to a luminosity of 8.89 $\times \rm 10^{37}~erg~s^{-1}$. Assuming no \Lya loss due to scattering or dust attenuation, the in-situ SFR would be $\rm 10^{-4}~M_{\odot}~yr^{-1}$ associated with the DLA \citep[as discussed by][]{wolfe05}. \citet{moller04} correlated the presence of \Lya emission to the metallicity of the system. They argue that since SFR and mass of galaxies are correlated and the mass and the metallicity of galaxies are correlated, hence one would except the \Lya production rate and metallicity to correlate as well. Interestingly, the metallicity of our DLA falls within the range of metallicities of DLAs in their sample with \Lya emission.

Another strong possibility for the origin of the \Lya emission feature was scattered radiation generated by young stars in galaxies~B and C. 
Such low levels of \Lya emission are consistent with extended \Lya-emitting halo around star-forming galaxies \citep[][and references therein]{wisotzki16, hayes13, lake15, kunth03, hayes05, ostlin14,mas_ribas16}. The COS aperture is $\approx$ 1.25$^{\prime\prime}$ in radius at a radial distance of 110$^{\prime\prime}$ and 190$^{\prime\prime}$ from galaxies B and C, respectively. If a uniform \Lya halo exists around galaxies B or C, the total \Lya luminosities would be 0.7$\times \rm 10^{42}erg~s^{-1}$ and 2$\times \rm 10^{42}~erg~s^{-1}$ respectively. These fluxes were similar to that measured by \citet{wisotzki16} in their sample of high-redshift galaxies on an individual object-by-object basis.
The \Lya luminosities would imply an SFR of at least 0.72 and 2.2~$\rm M_{\odot}~yr^{-1}$ for galaxies B and C respectively. Interestingly, both these values were slightly smaller than the measured SFRs (see Table~1). That was expected, as \Lya is attenuated in the presence of dust. Therefore, it is highly likely that these galaxies may indeed have an \Lya halo around them. Such \Lya halos in gas-rich groups and protoclusters had been detected at higher redshifts \citep[e.g. at z$\sim$2 by][]{prescott12}.

 \begin{figure*}
  \includegraphics[trim=0mm 0mm 110mm 0mm,  clip=true, scale=.65, angle=-90]{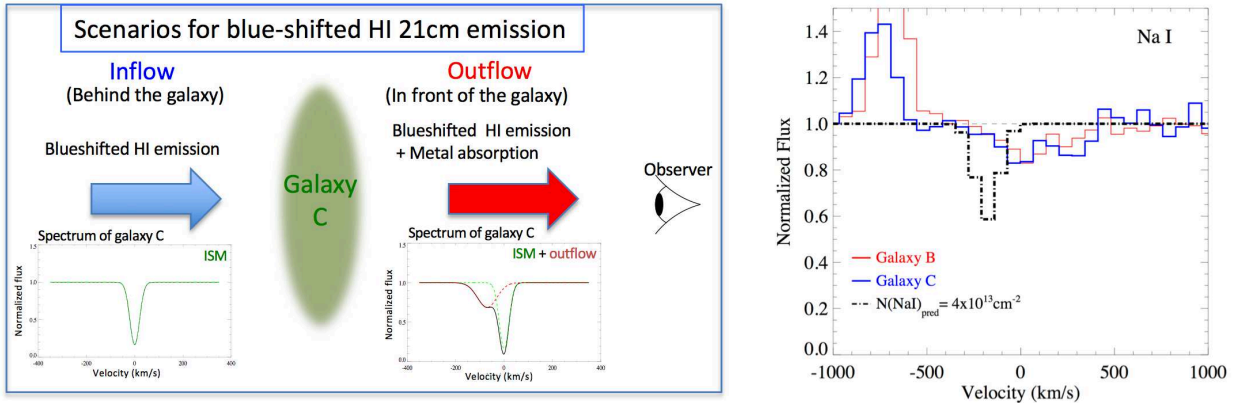}
 \caption{Left: cartoon describing two possible scenarios (inflow and outflow) that explain the possible origins of blueshifted  \textsc{Hi} emission in galaxy~C. In the case of an outflow capable of generating blue-shifted \textsc{Hi} emission, we expect strong  blueshifted metal-line absorption in the spectrum of galaxy~C. In the absence of any signs of a strong outflowing wind such as strong blue-shifted Na~I absorption or any  broad emission lines, we concluded that the most likely scenario for the origin of the blue-shifted \textsc{Hi} was gas inflow. Right: Sloan Digital Sky Survey spectra showing the Na~I$\lambda \lambda$~5890, 5896 doublet for galaxies B and C. We did not detect any outflowing component in the spectra of galaxy C where blueshifted \textsc{Hi} emission was detected. The predicted absorption profile of the blueshifted Na~I  with $\rm N(Na I)=4\times10^{13}~cm^{-2}$ is shown in black. The profile was assumed to have the same b-value as the bluemost component of \textsc{O~i} profile. The predicted Na~I column density was estimated assuming a solar abundance ratio of  $\rm N(Na)/N(O) = 3.8\times10^{-3}$. We assumed that all the Oxygen was in \textsc{O~i} and the sodium in Na~I. We also did not detect blueshifted Ca~II. }
 \label{inflow_cartoon}
\end{figure*}

\subsection{Inflowing Atomic Gas Detected in Galaxy C} 

The kinematics of the bluest component of \ion{O}{1} in the DLA matched well with the blueshifted 21~cm \HI emission toward galaxy~C.  Therefore, we concluded that the QSO sight line and the \HI emission near galaxy~C were tracing the same structure. This structure was most likely spatially distinct from the DLA and was just a chance overlap on the sky toward the QSO sight line.
The position-velocity diagram (Figure~\ref{HI_pos_vel_BC}) shows a velocity extension of 200~\kms of the \HI associated with galaxy C that was seen in both the east-to-west and north-to-south slices. The positional spread of the blueshifted \HI remained the same in the east-to-west direction, whereas it moved slightly toward the east (away from the center of the galaxy) in the north-to-south slice. The positional shift was not well resolved (synthesized beam size of $\rm 42^{\prime\prime}\times 35^{\prime\prime}$). The spatial distribution of blueshifted \HI indicated it to be concentrated at the center of galaxy~C and was not resolved in our most sensitive \HI map. 

The presence of the blueshifted \HI may be interpreted as either (1) inflow into the galaxy from behind or (2) outflow from the galaxy that was moving toward us (see the left panel of Figure~\ref{inflow_cartoon} for a schematic). 
The expected signatures of these scenarios were indistinguishable in the \HI 21~cm data if the gas was sufficiently neutral. However, other metal-line absorption in the spectrum of galaxy~C could be used to distinguish between them.
In the first scenario, we should not see any blueshifted absorption in the galaxy spectra, as the inflow would be behind the source, i.e. the stars in the galaxy. On the other hand, for outflowing gas, we should detect it as blueshifted absorption in the spectrum  of the galaxy.

The outflowing gas is expected to be metal-rich.  Since we detected neutral hydrogen, we expected the presence of low ionization metal lines such as \ion{Na}{1} and \ion{Ca}{2}. These observed wavelengths for these transitions were covered in the optical SDSS spectra of galaxy C (see right panel of Figure~\ref{inflow_cartoon}). 
We used the bluest component of the \ion{O}{1} absorber of the DLA, which matches the kinematics of the blueshifted \HI emission, to predict the column density of \ion{Na}{1} if the blueshifted \HI was outflowing. We assumed the sodium-to-oxygen ratio to be solar and derived a the predicted $\rm N(NaI)=4\times 10^{13}~cm^{-2} $. The predicted \ion{Na}{1} profile is shown in the right panel of Figure~\ref{inflow_cartoon}. This was calculated assuming that the Doppler b-parameter of \ion{Na}{1} is the same as that of the bluest component of \ion{O}{1}  (i.e. 41~\kms).

The SDSS spectrum did not show \ion{Na}{1} absorption at the predicted level. Instead, we estimated the limiting column density of $\rm N(NaI)_{3\sigma}<5\times 10^{12}~cm^{-2} $, which was an order of magnitude smaller than the predicted.
Hence, it is likely that the blueshifted \HI was not tracing outflowing gas; instead, it was tracing inflowing gas from behind the galaxy. Since \ion{Na}{1} may suffer from ionization effects and depletion, we suggest that this result is verified by the sensitive ultraviolet spectroscopy of galaxy~C tracing \Lya and other metal-line species.

Corroborative evidence supporting the proposed inflow scenario was that the ionization state of the gas in the blueshifted wing was fairly low with no \ion{Si}{4}. Unlike inflowing cold gas, outflows are driven by hot wind fluid sweeping the cold ISM. 
The outflows should therefore have signatures of iterations of cold clouds with the hot ($\rm 10^{6-7} ~K$) wind fluid. These signatures would include boundary layers of gas at various ionization states (traced by \ion{Si}{4}, \ion{C}{4}, and \ion{O}{6}) that envelop the cooler cores \citep[see, e.g.][]{sembach03,songaila06, grimes09, chisholm18}. We did not detected any boundary layers.

 We estimated the accretion rate of gas to be  $\sim \rm 2 M_{\odot}~yr^{-1}$. Using the observed column density of the bluemost component in the \ion{O}{1}  at -177~\kms and assuming its metallicity to be the same as the DLA, we estimated the total hydrogen column density of $\rm N(H) \ge 6.2\times 10^{19}~cm^{-2}$. The motivation for choosing \ion{O}{1} was due to its ionization potential matching hydrogen, as well as the fact that the line kinematics are clearly observed in its profile, unlike  the \Lya profile. 
We assumed the solar abundance ratio of O/H and that all the oxygen was in \ion{O}{1} to derive the above-mentioned value. The estimated \HI column density implies that the inflowing gas was a Lyman limit system (LLS). The LLSs have been argued to be tracers of inflow of gas in galaxies \citep[][and references therein]{hafen17,fumagalli16, erkal15, lehner12, lehner13, ford13, vandevoort12,fumagalli11}. 
Photoionization modeling suggested that the density of the LLS was in the range of $\rm 0.3-1.2\times 10^{-3}~cm^{-3}$. This implied a line-of-sight length of $\rm 16-64$~kpc. 
If the bulk of the gas was moving at 177~\kms toward galaxy C, then the accretion rate was $\rm 1.4-5.7~M_{\odot}~yr^{-1}$. 

 Based on the transverse size of the blueshifted gas and the halo mass of the group, we suspected that inflowing structure might be of tidal origin and was falling into the nuclear region of galaxy~C.
The strong H$\alpha$ emission in galaxy~C supported a recent episode of star formation. This made galaxy~C an excellent candidate for studying how gas inflow during interactions may trigger star formation and may lead to a starburst. The accretion rate derived from the \ion{O}{1} absorption measurements matched the SFR ($\rm 2.5~M_{\odot}~yr^{-1}$) of galaxy~C. Detailed study of the stellar population and its distribution will provide critical clues to connecting gas inflow to the physics of nuclear star formation.

\section{Summary  \label{sec:conclusion}} 

We presented a QSO sight line from HST program 12467 that probed the central regions of a galaxy group at z=0.029.  The ultraviolet spectrum was obtained with the high-resolution G130M grating of the COS and had a spectral resolution of 15-20~\kms. 

The group consisted of six spectroscopically confirmed members (galaxies A-F) with a stellar mass ranging from $\rm 10^{\sim9-11}~M_{\odot}$.
We detected a DLA system at the redshift of the group along with a large assortment of metal-line transitions. Motivated by this discovery, and in order to probe the nature and origin of the DLA, we imaged the group in the \HI 21cm hyperfine transition with the VLA in D  and C  configurations. Here we list the conclusions of our study and how this system may help in bridging the gap in our understanding of interaction-driven star formation.

\begin{itemize}
\item[1.] We detected a DLA with atomic hydrogen column density of N(HI)=10$\rm ^{20.4}~cm^{-2}$ at the redshift of the group. We also detected \Lya emission at the absorption trough of the DLA. The \Lya luminosity suggested that the emission could be associated with an \Lya halo surrounding galaxies B and/or C.

\item[2.] We also detected low ionization transitions such as \ion{O}{1}, \ion{C}{2}, \ion{Si}{2}, \ion{S}{2} and intermediate-ionization transitions such as \ion{Si}{3}. These transitions had multiple components spread over a range of velocities ($\approx$250\kms). The component structure varied between the species, thus suggesting that the DLA contained multiple components with slightly different ionization states. However, all the components are primarily neutral. The absence of high-ionization transitions such as \ion{N}{5} or \ion{Si}{4}, along with evidence from ionization modeling, confirmed this finding.

\item[3.] The metallicity of the gas was estimated to be $\rm 0.33^{+0.17}_{-0.11}~Z_{\odot}$. The estimate relied on the robust column density measurements of the \Lya transition and optically thin \ion{S}{2}. The metallicity of the DLA was consistent with that seen in outer disks of galaxies.

\item[4.] Photoionization modeling suggested  that the metal-line ratios seen in the DLA were consistent with the DLA system being ionized by the cosmic ultraviolet background. It also constrained the average densities of the DLA to be $\rm 2.5 \times10^{-3} ~particles~cm^{-3}$. This implied that the absorbing structure is at least 29~kpc in size along the line of sight.

\item[5.] The VLA HI image of the group showed a gas-rich environment with atomic gas associated with galaxies~B, C, E, and F. The \HI mass of the group was estimated to be $\rm 2\times 10^{10}~M_{\odot}$  with most of it concentrated as an envelope surrounding galaxies~B and C.

\item[6.]No \HI counterpart to the DLA was seen in the VLA image. We found finger-like extensions of the \HI associated with galaxies B and C extending toward the QSO sight line. Based on the proximity of the \HI complex enveloping galaxies B$+$C to the QSO sight line and the metallicity of the DLA, we concluded that the DLA was most likely tracing gas once part of the \HI complex. The morphology of the B$+$C complex suggested that it might have been shaped by mild tidal forces. We also evaluated the possibility of dwarf galaxy as the origin of the DLA. Based on the limiting \HI mass, stellar mass, and SFR surface density, the DLA is unlikely to be tracing the ISM of a dwarf galaxy. However, this possibility cannot be entirely ruled out. 

\item[7.] We discovered blueshifted \HI (200~\kms) associated with galaxy C. The blueshifted gas had $\rm M(HI) = 2.2 \times 10^{9}~M_{\odot}$. This component was concentrated around galaxy C and was unresolved in the VLA image. We concluded that this was tracing gas flowing into galaxy C from behind. The alternate explanation was outflowing gas, which was rejected owing to the absence of metal-line (\ion{Na}{1}) absorption in the optical spectra of the galaxy. The galaxy showed very strong H$\alpha$ emission, thus indicating the presence of young stars of ages less than 5~Myrs. Therefore, this galaxy is a unique example of gas accretion and star formation going hand in hand. 

\item[8.]  The bluemost component of the \ion{O}{1} profile of the DLA matched well with the blueshifted emission in the \HI 21~cm profile of galaxy~C and the filamentary structure around galaxy~E. This suggested that the blueshifted gas may be a group-wide structure. We estimated the gas accretion rate of $\rm 2~M_{\odot}~yr{-1}$ for galaxy~C, which is matches its SFR.

\end{itemize}

Galaxies B and C are excellent candidates for studying how gas inflow during interactions may trigger star formation. 
The presence of strong H$\alpha$ emission in galaxies B and C as well as the likely detection of a Ly$\alpha$ halo, made this system a unique laboratory to study inflow and outflow processes. 
The nature of the final product of the interaction of these galaxies is hard to predict. However, the discovery of possible evidence of cold-gas accretion and the formation of young H$\alpha$-producing stars may provide us with valuable insights into the processes leading to a starburst. In the future, detailed study of the stellar populations, the ISM, and other aspects of this group will shed light on the triggers for interaction-driven gas flows and perhaps even what triggers starbursts.

\vspace{.5cm}
\acknowledgements 
We especially thank the referee for his/her constructive comments.
We thank the support staff at the STScI, the GBT, and the VLA for help and support during the observations.
We also thank  Xavier Prochaska, Mary Putman, Cameron Hummels, Jessica Werk, Marcel Neelman, and Rongmon Bordoloi for extensive (and useful) discussions.  
This work is based on observations with the NASA/ESA Hubble Space Telescope, which is operated by the Association of Universities for Research in Astronomy, Inc., under NASA contract NAS5-26555. SB and TH were supported by grant HST GO 12603.
BC is the recipient of an Australian Research Council Future Fellowship (FT120100660). 
This project also made use of SDSS data. Funding for SDSS-III has been provided by the Alfred P. Sloan Foundation, the Participating Institutions, the National Science Foundation, and the U.S. Department of Energy Office of Science. The SDSS-III web site is http://www.sdss3.org/.
SDSS-III is managed by the Astrophysical Research Consortium for the Participating Institutions of the SDSS-III Collaboration including the University of Arizona, the Brazilian Participation Group, Brookhaven National Laboratory, Carnegie Mellon University, University of Florida, the French Participation Group, the German Participation Group, Harvard University, the Instituto de Astrofisica de Canarias, the Michigan State/Notre Dame/JINA Participation Group, Johns Hopkins University, Lawrence Berkeley National Laboratory, Max Planck Institute for Astrophysics, Max Planck Institute for Extraterrestrial Physics, New Mexico State University, New York University, Ohio State University, Pennsylvania State University, University of Portsmouth, Princeton University, the Spanish Participation Group, University of Tokyo, University of Utah, Vanderbilt University, University of Virginia, University of Washington, and Yale University.

{\it Facilities:} \facility{COS ()} \facility{VLA ()} \facility{GBT ()} \facility{Arecibo ()} \facility{Sloan ()}

\bibliographystyle{apj}	        
\bibliography{myref_bibtex}		

\clearpage
\LongTables 
\begin{landscape}
\begin{deluxetable}{clccc ccccc   ccccc ccccc}  
\tabletypesize{\scriptsize}
\tablecaption{Properties of Member Galaxies in the Group.\label{tbl-sight lines}}
\tablewidth{0pt}
\tablehead{
\colhead{Label} & \colhead{Galaxy} & \colhead{R.A.} &  \colhead{Decl.} & \colhead{z}   & \colhead{log M$_*^{a}$}    & \colhead{log~SFR$^{a}$}   & \colhead{log~sSFR$^{a}$}&  \colhead{O/H$^a$}   &   \colhead{Color$^b$}   &   \colhead{Impact Param.} \\
\colhead{}    & \colhead{}    &   \colhead{}             & \colhead{}                & \colhead{}              & \colhead{($\rm log~M_{\odot}$)}    & \colhead{($\rm log~M_{\odot}~yr^{-1}$)}        & \colhead{($\rm log~yr^{-1}$)}   & \colhead{}    &   \colhead{}  &   \colhead{(kpc)}  }
\startdata
A &  J151243.69$+$012753.4$^c$  & 228.182 & 1.46451 &  0.0293 &  10.92 &   0.70 & -10.2 & $-$   & Blue & 65\\
B &  J151233.08$+$013017.3          & 228.137 & 1.50449 &  0.0292 &    9.55 &   0.20 &  -9.3 &  8.73 & Blue & 64 \\
C &  J151225.15$+$012950.4          & 228.104 & 1.49705 &  0.0289 &    9.81 &   0.40 &  -9.4 &  8.73   & Blue & 112 \\
D &  J151249.84$+$012827.7          & 228.207 & 1.47404 &  0.0304 &    9.05 &  -0.40 &  -9.4 &  8.59   & Blue & 112 \\
E &  J151207.62$+$013123.8          & 228.031 & 1.52303 &  0.0284 &    9.98 &   0.15 &  -9.8 &  9.03   & Blue & 276 \\
F &  J151223.37$+$013823.9           & 228.097  & 1.63999 &  0.0285 &    9.64 &   -0.05 &  -9.6 &  9.01   & Blue & 360 
\enddata
\tablenotetext{a}{M$_*$, SFR, sSFR, and metallicities are adopted from MPA-JHU catalog DR4. M$_*$ is the total mass corrected for the fiber placement.}
\tablenotetext{b}{Color and luminosity are derived from NYU-VAGC.}
\tablenotetext{c}{Galaxy part of the COS-GASS survey.}
\end{deluxetable}
\clearpage
\end{landscape}

\clearpage
\LongTables 
\begin{landscape}
\begin{deluxetable}{llccccccccc}
\tabletypesize{\scriptsize}
\tablecaption{Transitions Associated with the Damped \Lya System.  \label{tbl-J085301_J092844}}
\tablewidth{0pt}
\tablehead{
 \colhead{Transition} &\colhead{$\lambda_{rest}$} & \colhead{$\rm W_{rest}^a$}  & \colhead{Centroid $^b$} &\colhead{b$^b$}                    &\colhead{log~N}  \\
 \colhead{}                 & \colhead{$\rm (\AA)$}       & \colhead{$\rm (m\AA)$}   &\colhead{($\rm km~s^{-1}$)} &\colhead{($\rm km~s^{-1}$)} &\colhead{($\rm log~ cm^{-2}$)} }
\startdata
 
  \textsc{H~i} & 1215.67$^b$  & 10553$\pm$190   &       44$\pm$13 &   81$^{+270}_{-63}$ &  20.42$\pm$0.10  \\
  \\
  \textsc{O~i}         & 1302.17$^d$   &  995$\pm$40   & 27$\pm$2, -177$\pm$24, -99$\pm$6 &
41$^{+7}_{-6}$,73$^{+43}_{-27}$, 21$^{+16}_{-9}$  &    16.0$\pm$0.4, 16.0$\pm$0.2, 14.3$\pm$0.3 \\

 \\
\textsc{Si~ii}           & 1190.42$^e$  &      429$\pm$14 & \multirow{4}{*}{$\Big] \scriptsize \rm 70\pm2, -7\pm3, -139\pm4, -215\pm4$}  & \multirow{4}{*}{$\Big] \scriptsize \rm $25$^{+2}_{-2}$, 39$^{+5}_{-4}$, 49$^{+7}_{-6}$, 17$^{+8}_{-6}$}  & \multirow{4}{*}{$\Big]\scriptsize>13.97, >13.70, >13.59,12.83\pm0.16$ }  \\  
        & 1193.29$^e$   &     677$\pm$23   \\
         & 1260.42$^d$   &  1040$\pm$47  &&&  \\
         & 1304.37   &    336$\pm$32  \\
 \\
\textsc{S~ii}              & 1250.58   &      73$^f$    & \multirow{3}{*}{$\Big] \scriptsize \rm 63 \pm 8$}   & \multirow{3}{*}{$\Big]\scriptsize\rm16^{+50}_{-12}$}  & \multirow{3}{*}{$\Big ] \scriptsize 15.1\pm0.2 $} \\  
       & 1253.81   &     117$\pm$22   \\
          & 1259.52   &     147$^f$   \\
 \\
 \textsc{Si~iii}              & 1206.50   &  678$\pm$60   & 22$\pm$6, -165$\pm$5      &   73$^{+10}_{-9}$, 55$^{+8}_{-7}$      &      13.4$\pm$0.04, 13.3$\pm$0.05 \\
 \\
 \textsc{Si~iv}            & 1393.76   &  $<$300$^g$   & -  & - & $<$13.5$^g$ \\ 
\\
 \textsc{C~ii}              & 1334.53$^c$   &  1586$\pm$70   & 21$\pm$4, -130$\pm$8, -278$\pm$11 &  46$^{+11}_{-9}$, 49$^{+16}_{-12}$, 81$^{+15}_{-13}$ &  $>$15.4, 14.4$\pm$0.1, 14.6$\pm$0.1  \\
 
\enddata
\tablenotetext{a}{The error takes into account both the statistical error and the error in continuum fitting.}
\tablenotetext{b}{The HI measurement was also included by \citet{N16} in their archival study of DLA systems.}
\tablenotetext{c}{The centroids and Doppler b-values were obtained after correcting for the appropriate line-spread function of grating G130M. The velocities are with respect to the group rest frame corresponding to z=0.029268.}
\tablenotetext{d}{Saturated lines.}
\tablenotetext{e}{Partly saturated lines.}
\tablenotetext{f}{Based on the Voigt profile fit to S~II 1253.}
\tablenotetext{g}{Equivalent width and column density correspond to 3σ uncertainty in measurements at the position of the expected line. The column density was derived assuming the transitions to be optically thin.}
\end{deluxetable}
\clearpage
\end{landscape}

 \clearpage

 \appendix
 
 \subsection{Appendix: Single-dish \HI spectra}

The GBT data were obtained under program GBT-14A-377. We used the dual-polarization L-band system with two intermediate frequency (IF) modes and nine-level sampling. The IFs were set to yield a channel width of 1.56 kHz (0.33~\kms) using 8192 channels over a total bandwidth of 12.5 MHz. Observations were made in the standard position-switching ON-OFF scheme with 300~s at each position. Data were recorded at 10~s intervals to minimize the effect of radio frequency interference.
The ON position was chosen to the position of the QSO sight line (unlike galaxy A for the Arecibo observations), and the OFF position was chosen to be +20$^{\prime}$ offset in Right Ascension from the ON position. 
This was done to track the possible presence of a source in the OFF position. We used standard flux calibrators for pointing and estimating antenna gain. Local pointing corrections (LPCs) were performed using the observing procedure AutoPeak and the corrections were then automatically applied to the data. We expect this to result in a pointing accuracy of 3$^{\prime\prime}$. Similarly, flux calibrations were performed by applying antenna gain to data for each session separately. The corrections also include attenuation due to the opacity of the air (around 1\%). The data were analyzed using the NRAO software GBTIDL.

\begin{figure*}[!h]
 
 \hspace{1.0cm}
\includegraphics[trim=0mm 0mm 0mm 0mm,  clip=true, scale=0.45]{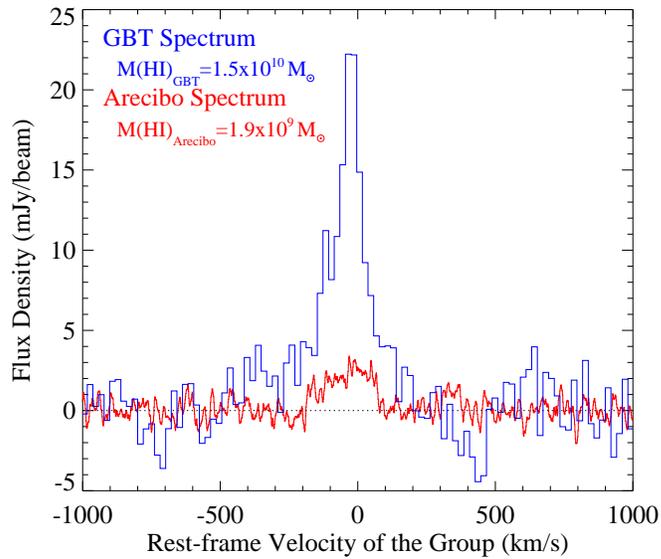}
 \caption{The \textsc{Hi} spectrum toward the group as seen by the GBT and the Arecibo Telescope. The Arecibo spectrum was obtained as part of the GASS survey. The GBT spectrum shows almost an order of magnitude more \textsc{Hi}  than that seen in the Arecibo spectrum. The \textsc{Hi} mass estimated from the GBT and the Arecibo spectra are $\rm 1.5\times 10^{10}$ and $\rm 1.9\times 10^{9}~M_{\odot}$ respectively. This suggests that most of the HI are located in the region away from the Arecibo beam coverage. The location of the \textsc{Hi} seen in the Arecibo spectrum is near the two star-forming - galaxies B and C. We have confirmed that with our VLA \textsc{Hi} mapping. The \textsc{Hi} strength is highly attenuated in the Arecibo spectrum due to the offset in the beam positioning with respect to the \textsc{Hi} distribution. }
 \label{HI_spectra}
\end{figure*}

Figure~\ref{HI_spectra} plots the GBT spectrum along with the Arecibo telescope spectrum. The total \HI mass associated with the group as estimated from the GBT spectrum is $1.5\times 10^{10}~M_{\odot}$ with an  uncertainty of at most 10\% due to calibration uncertainties. 
The GBT \HI mass almost an order of magnitude more than the \HI detected in Arecibo spectrum. 
This indicates that a large fraction of the \HI associated with the group resides outside the Arecibo beam coverage. The GBT beam (FWHM of 9.1$^{\prime}$), being almost 7 times larger in area than the  Arecibo beam (FWHM of 3.5$^{\prime}$), covers a large part of the galaxy group. Furthermore, the GBT beam was centered at the position of the QSO whereas the Arecibo beam was centered at galaxy~A. Therefore, we concluded that most of the \HI in the group is not associated with the galaxy A but was distributed in the region covered by the GBT beam and not the Arecibo beam (see Figure~1).

The spectral profile of the \HI distribution seen by the GBT is similar to that detected by the VLA in \HI complex enveloping galaxies B$+$C; however, the total mass is slightly lower. That is expected; as the GBT pointing was $\approx \rm4^{\prime}$ from the center of the \HI distribution, which corresponds to a drop of 50\% in the primary beam sensitivity. We confirmed the consistency of the single-dish date and VLA data by overlaying the beam shape on the VLA image using a method similar to the analysis by \citet[][]{borthakur15b}.

\end{document}